\documentclass[aps,showpacs,twocolumn,floatfix,prd]{revtex4}
\usepackage{epsfig,graphics,color}
\usepackage{graphicx}% Include figure files
\usepackage{dcolumn}% Align table columns on decimal point
\usepackage{bm}% bold math
\usepackage{amsmath}
\usepackage{eucal,eufrak}

\newcommand \tie {{\it i.e.}}

\newcommand \f {\not\!}
\newcommand \fs {\not}

\newcommand \kd  {\delta}
\newcommand \ra  {\rightarrow}

\newcommand \fp {{\bf p}}

\newcommand \fy {{\bf y}}

\newcommand \h {\theta}

\newcommand \vl {\vec{l}}

\newcommand \vp {\vec{p}}

\newcommand \g {\gamma}
\newcommand{\gm}{\gamma^-}
\newcommand{\gt}{\gamma^{\perp}}
\newcommand{\gp}{\gamma^+}

\newcommand \e {\epsilon}

\newcommand \p {{\prime}}

\newcommand \x {\cdot}
\newcommand \hf {\frac{1}{2}}
\newcommand \A {\alpha}
\newcommand \B {\beta}

\newcommand \lc {\langle}
\newcommand \rc {\rangle}
\newcommand \prt {\partial}

\newcommand \sg {\sigma}

\newcommand \nt {\noindent}

\newcommand \md {\mathfrak{D}}

\newcommand \bvec{\left( \begin{array}{c} }
\newcommand \evec{\end{array} \right)}
\newcommand \tr {\mbox{{\bf Tr}}}
\newcommand \eg {{\it e.g.}}  
\newcommand \bea{\begin{eqnarray} }
\newcommand \eea{\end{eqnarray} }
\newcommand \nn {\nonumber}
\newcommand \be {\begin{equation}}
\newcommand \ee {\end{equation}}

\newcommand \mbx {\mbox{}}

\newcommand \psibar {\bar{\psi}}
\newcommand \ata {& \times &}

\newcommand \slm {\sum\limits}
\begin{document}

\title{Higher twist jet broadening and classical propagation}

\author{A. Majumder}
\affiliation{Department of Physics, Duke University, Durham, NC 27708, USA}

\author{B. M\"uller}
\affiliation{Department of Physics, Duke University, Durham, NC 27708, USA}

\date{ \today}

\begin{abstract} 
The transverse broadening of jets produced in deep-inelastic scattering (DIS) off a large 
nucleus is studied in the collinear limit. A class of medium enhanced higher twist 
corrections are re-summed to calculate the transverse momentum 
distribution of the produced collinear jet. In contrast to previous approaches, re-summation 
of the leading length enhanced higher twist corrections is shown to lead to a two dimensional diffusion 
equation for the transverse momentum of the propagating  jet. Results for the average transverse 
momentum obtained from this approach are then compared to the broadening 
expected from a classical Langevin analysis for the propagation of the jet under the 
action of the fluctuating color Lorentz force inside the nucleons. The set of approximations 
that lead to identical results from the two approaches are outlined. The relationship between 
the momentum diffusion constant $D$ and the transport coefficient $\hat{q}$ is explicitly 
derived.
\end{abstract}


\maketitle

%%%%%%%%%%%%%%%%%%%%%%%%%%%%%%%%%%%%%%%%%%%%%%%%%%%%%%%%%%
%%%%%%%%%%%%%%%%%%%%%%%%%%%%%%%%%%%%%%%%%%%%%%%%%%%%%%%%%%
%%%%%%%%%%%%%%%%%%%%%%%%%%%%%%%%%%%%%%%%%%%%%%%%%%%%%%%%%%
%%%%%%%%%%%%%%%%%%%%%%%%%%%%%%%%%%%%%%%%%%%%%%%%%%%%%%%%%%
%%%%%%%%%%%%%%%%%%%%%%%%%%%%%%%%%%%%%%%%%%%%%%%%%%%%%%%%%%

 \section{introduction}

%%%%%%%%%%%%%%%%%%%%%%%%%%%%%%%%%%%%%%%%%%%%%%%%%%%%%%%%%%
%%%%%%%%%%%%%%%%%%%%%%%%%%%%%%%%%%%%%%%%%%%%%%%%%%%%%%%%%%
%%%%%%%%%%%%%%%%%%%%%%%%%%%%%%%%%%%%%%%%%%%%%%%%%%%%%%%%%%
%%%%%%%%%%%%%%%%%%%%%%%%%%%%%%%%%%%%%%%%%%%%%%%%%%%%%%%%%%

The study of dense matter through its effect on hard partonic jets is now 
an established science. Experiments at the Relativistic Heavy-Ion collider (RHIC) and 
the upcoming Large Hadron Collider (LHC) have considerable resources 
dedicated to the study of the modification of  high transverse momentum particles 
produced in the fragmentation of partonic jets which originated within and propagated 
through the produced dense matter~\cite{white_papers}. There currently exists a large 
number of theoretical models which provide quantitative estimates of the observed 
data~\cite{quenching,AMY,HT,GLV,ASW}.

While the different models have rather diverse origins, within the approximations made, 
similarities are pervasive. A well know example is the quantitative similarity between the schemes of 
Refs.~\cite{GLV} and~\cite{ASW} as demonstrated in Ref.~\cite{Salgado:2003gb}. 
The object of this article is two fold: primarily, the aim is a re-derivation of the transverse broadening
of a hard jet brought about by its propagation in dense matter with fluctuating color fields. 
Secondly, this represents the first effort  to find a physical resemblance between the 
Higher-Twist and Finite temperature field theory jet-quenching schemes  
Refs.~\cite{AMY,HT}. The focus will lie predominantly on the Higher-Twist approach 
of Ref.~\cite{HT}. In this scheme of energy loss, one tends to re-sum a class of diagrams 
which encapsulate the leading medium length enhanced power corrections to the basic 
process of a hard parton radiating a soft collinear gluon. Such gluon showers, emanating 
from hard partons in vacuum, often referred to as vacuum energy loss~\cite{GLV} (or zeroth-order 
energy loss~\cite{Djordjevic:2003zk}), lead to the DGLAP evolution~\cite{gri72} of parton-hadron 
fragmentation and structure functions. Such showers, in a medium, are influenced by the 
time (or length) dependent transverse momentum fluctuations experienced by the hard 
propagating jet and its radiated gluons and lead to a loss of the energy of the propagating 
parton. The effect of the transverse ``kicks'' from the medium on the systematics of energy loss 
may be divided into two parts. The effect on the propagation of the jet as well as its radiated gluons and 
the effect on the radiation vertex. 
The current work, intends to focus exclusively on the former \tie, the effect of the color 
fields of the medium on 
the transverse momentum fluctuations or 
\emph{broadening} experienced by a single hard parton as it traverses the medium. Radiation and 
energy loss will be dealt with in a future effort.

It is well known that hard partons, traversing dense matter, tend to 
pick up a transverse momentum which depends on the length of the medium~\cite{lqs}. The calculation 
of this length dependent  broadening is  achieved via the re-summation 
of a class of higher twist diagrams~\cite{Fries:2002mu} which are enhanced by the length traversed by the 
jet in  the medium. Within the assumptions made in such a calculation, the broadening may also be 
obtained from a purely classical analysis of a 
hard charged particle moving under the influence of a color Lorentz force. In this sense, there
exists a physical similarity with the basic picture underlying the jet quenching scheme of 
Ref.~\cite{AMY} which also admits a kinetic theory description of hard partons moving 
under the influence of soft fields. 

However, the existing literature on the inclusion of power corrections to energy loss or jet broadening processes 
has yet to yield such a simple physical picture. This will be the object of the current article.
This is hardly the first attempt to identify and re-sum the class of higher twist corrections required for the 
computation of jet broadening in extended media. A thorough derivation is provided to 
elucidate the basic assumptions and approximations made. 
Similar to previous derivations, the original parton emanating from the hard scattering is assumed to 
posses a vanishing transverse momentum distribution ($\delta^2({ \vec{p}}^{\,\,0}_\perp)$). 
The derivation is carried out in the high energy limit; as a result, the coupling of the hard parton 
with the medium is assumed to be weak. A derivation of the transverse broadening, when the 
parton couples strongly with the 
medium (albeit for a heavy quark) has been presented in Ref.~\cite{Casalderrey-Solana:2007qw}.
In contrast to previous derivations, re-summation of the leading higher twist corrections, leads to 
a diffusion equation for the transverse momentum distribution of the final parton on exit 
from the dense medium. The diffusion equation is solved and the relationship between the 
diffusion tensor and the energy loss transport coefficient $\hat{q}$ derived. 
The computation of the first non-vanishing moment  of transverse momentum distribution $\lc p_T^2 \rc$ 
is then re-derived within a classical approximation where a colored charge moves under the influence of 
the fluctuating Lorentz force within nucleons. This calculation is carried out using a simple Langevin analysis.  
The similarities between the two results and its implications are discussed.

The paper is organized as follows:
in the next section, the leading twist parton distributions will be factorized at leading order. In Sect.~III, 
the class of higher twist diagrams which are length enhanced at a given order $m$ will be identified 
and the leading contributions to the hadronic tensor will be calculated. 
In Sec.~IV, the hadronic tensor will be factorized into hard and soft piece and the $m^{\rm{th}}$ transverse 
momentum derivative of the hard part computed. In Sec.~V, an all order re-summation is carried out and the 
transverse momentum diffusion equation is derived. The diffusion equation will be 
solved and the moments of the transverse momentum distribution computed and related with 
the energy loss parameter $\hat{q}$. In Sect.~VI, the Langevin analysis is carried out.  
Concluding discussions  and future directions will be presented in Sect.~VII.

%%%%%%%%%%%%%%%%%%%%%%%%%%%%%%%%%%%%%%%%%%%%%%%%%%%%%%%%%%
%%%%%%%%%%%%%%%%%%%%%%%%%%%%%%%%%%%%%%%%%%%%%%%%%%%%%%%%%%
%%%%%%%%%%%%%%%%%%%%%%%%%%%%%%%%%%%%%%%%%%%%%%%%%%%%%%%%%%
%%%%%%%%%%%%%%%%%%%%%%%%%%%%%%%%%%%%%%%%%%%%%%%%%%%%%%%%%%
%%%%%%%%%%%%%%%%%%%%%%%%%%%%%%%%%%%%%%%%%%%%%%%%%%%%%%%%%%

 \section{Leading  twist  and parton distribution functions}

%%%%%%%%%%%%%%%%%%%%%%%%%%%%%%%%%%%%%%%%%%%%%%%%%%%%%%%%%%
%%%%%%%%%%%%%%%%%%%%%%%%%%%%%%%%%%%%%%%%%%%%%%%%%%%%%%%%%%
%%%%%%%%%%%%%%%%%%%%%%%%%%%%%%%%%%%%%%%%%%%%%%%%%%%%%%%%%%
%%%%%%%%%%%%%%%%%%%%%%%%%%%%%%%%%%%%%%%%%%%%%%%%%%%%%%%%%%

The focus of this article is restricted to the semi-inclusive 
process of DIS off a nucleus in the Breit frame where  one jet with a transverse 
momentum $l_\perp$ is produced,  

\bea
e(L_1) + A(p) \longrightarrow e(L_2) + J(l_\perp) + X .
\label{chemical_eqn}
\eea

\nt
In the above equation, $L_1$ and $L_2$ represent the momentum of the 
incoming and outgoing leptons. The incoming nucleus of atomic mass 
$A$ is endowed with a momentum $Ap$. In the final state,
all hadrons ($h_1,h_2,...$) with momenta $p_1,p_2,\ldots$ are detected and their
momentum summed to obtain the jet momentum and $X$ denotes that the 
process is semi-inclusive.

\begin{figure}[htbp]
%\begin{center}
%  \epsfxsize 80mm
%\hspace{0cm}
\resizebox{3in}{1.5in}{\includegraphics[0in,0in][8in,4in]{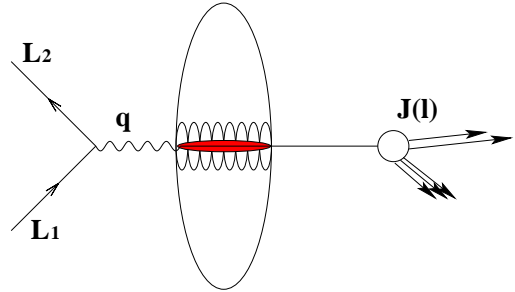}} 
%\vspace{0.25cm}
    \caption{ The Lorentz frame of the process where a nucleon in a large nucleus is
    struck by a hard space-like photon.}
    \label{fig1}
%  \end{center}
\end{figure}

The kinematics is defined in the Breit frame as sketched in Fig.~\ref{fig1}. 
In such a frame, the virtual photon $\g^*$ and the nucleus have 
momentum four vectors $q,P_A$ given as, 

\[
q = L_2 - L_1 \equiv \left[\frac{-Q^2}{2q^-}, q^-, 0, 0\right], 
\mbox{\hspace{1cm}}
P_A \equiv A[p^+,0,0,0].
\]

\nt
In this frame, the Bjorken variable is obtained as 
$x_B = Q^2/2p^+q^-$. 

The differential cross section of the semi inclusive process with a jet with transverse 
momentum $l_\perp$  and four-momentum $l$ may be expressed as 

\bea
\frac{E_{L_2} d \sigma } {d^3 L_2 d^2 l_\perp } &=&
\frac{\A_{EM}^2}{2\pi s  Q^4}  L_{\mu \nu}  
\frac{d W^{\mu \nu}}{d^2 l_\perp}, \label{LO_cross}
\eea

\nt
where $s = (p+L_1)^2$ is the total invariant mass of the lepton nucleon 
system. The reader may have already surmised the form of the leptonic tensor 
as,

\bea 
L_{\mu \nu} = \frac{1}{2} \tr [ \f L_1 \g_{\mu} \f L_2 \g_{\nu}].
\eea

\nt
In the notation used in this paper, $| A; p \rc$ represents the initial state of 
an incoming nucleus with $A$ nucleons with a momentum $p$ per nucleon. The 
general final hadronic or partonic state is defined as 
$| X \rc $.
As a result, the semi-inclusive hadronic tensor may be defined as

\bea 
W^{\mu \nu}\!\!\!\!&=& \!\!\!\! \sum_X \!\!(2\pi^4) 
\kd^4 (q\!+\!P_A\!-\!p_X ) \nn \\
\ata \lc A; p |  J^{\mu}(0) | X  \rc \lc X  | J^{\nu}(0) | A;p \rc  \nn \\
&=& 2 \mbox{Im} \left[  \int d^4 y e^{i q \cdot y } \lc A;p | J^{\mu} (y) J^{\nu}(0) | A;p \rc \right]
\eea

\nt
where the sum ($\sum_X$)  runs over all possible hadronic states and $J^{\mu}$ is the 
hadronic electromagnetic current ($J^{\mu} =  Q_q \bar{\psi}_q \g^\mu \psi_q$, where $Q_q$ is the 
charge of a quark of flavor $q$ in units of the positron charge $e$). 
It is understood that the quark operators are written in the  interaction picture, and 
factors of the electromagnetic coupling constant have already been extracted and 
included in Eq.~\eqref{LO_cross}.
The leptonic tensor will not be discussed further. The focus in the remaining 
shall lie exclusively on the hadronic tensor. This tensor will be expanded 
order by order in a partonic basis and leading twist and maximally length enhanced higher 
twist contributions will be isolated.

\begin{figure}[htbp]
%\begin{center}
%  \epsfxsize 80mm
%\hspace{0cm}
  \resizebox{3.2in}{1.6in}{\includegraphics[0in,0in][8in,4in]{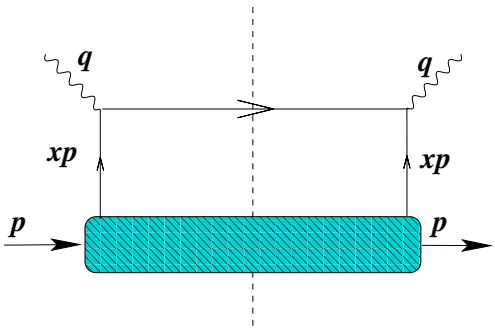}} 
%\vspace{0.25cm}
    \caption{ The Lowest order and leading twist contribution to $W^{\mu \nu}$.}
    \label{fig2}
%  \end{center}
\end{figure}

The leading twist contribution is obtained by expanding the products of 
currents at leading 
order. This contribution may be expressed diagrammatically 
as Fig.~\ref{fig2} and quantitatively 
expressed as (we also take the average over initial states and 
sum over final states to obtain)

\bea
{W_0^A}^{\mu \nu} 
% &\equiv & 2 Im \left[ \int d^4y e^{iq \x y} \lnuc  J^{\mu}(y) J^\nu (0)  \rnuc \right] \nn \\
%
&=&  C_p^A W_0^{\mu \nu} \label{w_mu_nu_twist=2}\\
%
%\tr \left[ \lnuc \psi(0) \psibar(y) \rnuc  \g^{\mu} (\f p + \f q) 2 \pi \kd [(p+q)^2]  \rnuc \right]
%
&=& C_p^A  \frac{2 \pi x_B}{2 Q^2} \tr \left[ \f{p} \g^\mu \left( \f{q} + x_B \f{p} \right) \g^\nu  \right] 
\sum_q Q_q^2f_q (x_B)  \nn \\
&=&C_p^A 2 \pi  [ g^{\mu -} g^{\nu +} + g^{\mu +} g^{\nu -} - g^{\mu \nu} ]  \sum_q Q_q^2 \nn \\
\ata \int \frac{d y^-}{2\pi} e^{-ix_B p^+ y^-} \hf \lc p| \psibar(y^-) \g^+ \psi(0) | p \rc \nn
\eea

\nt
In the above equation, $C_p^A$ expresses the probability to find a nucleon state inside a nucleus with $A$ nucleons and 
$f_q(x_B)$ is the quark structure function within a single nucleon state, this represents the expectation value of a 
twist 2 operator: the quark number operator.  
Throughout this paper, the light-cone component notation for four vectors ($p \equiv [p^+,p^-, \vec{p}_\perp]$) 
will be used, where,

\bea
p^+ = \frac{p^0 + p_z}{2}; \,\,\, p^- = p^0 - p_z.
\eea
\nt
In the collinear 
limit, the incoming parton is assumed to be endowed with very high forward momentum $({p_0}^+ = x_0 p^+, p_0^- \ra 0)$
with negligible transverse momentum ${p_0}_\perp \ll p_0^+$.  Within the 
kinematics chosen, the photon also has no transverse 
momentum. As a result, the produced final state parton also has a vanishingly small transverse momentum 
(\tie, with a distribution $\kd^2(p_\perp^2)$). 
As a result, the transverse momentum distribution of the produced parton is obtained as 
the differential hadronic tensor,

\bea
\frac{d^2 W_0^{\mu \nu}}{d^2l_\perp } = C^A_p W_0^{\mu \nu} \kd^2 (\vec{l}_\perp  ) \label{d_W_0}
\eea

It should be pointed out that in DIS on a nucleon, the final transverse momentum distribution of the 
produced jet is never a strict  $\kd$-function but rather a Gaussian with a width that depends on the 
scale and $x_B$ of the scattering. Neither is the final outgoing quark strictly on shell as the cut line 
in Fig.~\ref{fig2} would indicate. The outgoing quark will eventually fragment into a jet of hadrons 
and thus has a positive virtuality $m^2$ which is dependent on the scale $Q^2$ of the hard scattering. 
In a high energy scattering event $\Lambda_{QCD} << m < Q$, which allows futher re-scatterings of the 
quark to be treated perturbatively, yet $m << q^-$ which leads to its identification as a single jet. The 
scale $m$ may also be used to define the scale of the final fragmentation function. In the remaining 
sections, both the virtuality of the produced quark and its initial transverse momentum distribution will be 
taken to be vanishingly small. These approximations are carried out in the interest of simplicity and 
both quantities should be understood to be present.

%%%%%%%%%%%%%%%%%%%%%%%%%%%%%%%%%%%%%%%%%%%%%%%%%%%%%%%%%%
%%%%%%%%%%%%%%%%%%%%%%%%%%%%%%%%%%%%%%%%%%%%%%%%%%%%%%%%%%
%%%%%%%%%%%%%%%%%%%%%%%%%%%%%%%%%%%%%%%%%%%%%%%%%%%%%%%%%%
%%%%%%%%%%%%%%%%%%%%%%%%%%%%%%%%%%%%%%%%%%%%%%%%%%%%%%%%%%
%%%%%%%%%%%%%%%%%%%%%%%%%%%%%%%%%%%%%%%%%%%%%%%%%%%%%%%%%%

 \section{Higher twist and transverse broadening}

%%%%%%%%%%%%%%%%%%%%%%%%%%%%%%%%%%%%%%%%%%%%%%%%%%%%%%%%%%
%%%%%%%%%%%%%%%%%%%%%%%%%%%%%%%%%%%%%%%%%%%%%%%%%%%%%%%%%%
%%%%%%%%%%%%%%%%%%%%%%%%%%%%%%%%%%%%%%%%%%%%%%%%%%%%%%%%%%
%%%%%%%%%%%%%%%%%%%%%%%%%%%%%%%%%%%%%%%%%%%%%%%%%%%%%%%%%%

\begin{figure}[htbp]
%\begin{center}
%  \epsfxsize 80mm
%\hspace{0cm}
  \resizebox{2in}{2in}{\includegraphics[1.5in,0in][5.5in,4in]{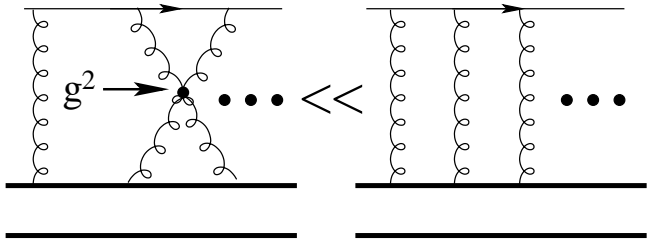}} 
%\vspace{0.25cm}
    \caption{ A higher order contribution suppressed by $g^2$ due to four gluon coupling.}
    \label{four_gluon}
%  \end{center}
\end{figure}

Higher twist 
contributions are obtained from diagrams  which  include expectation values of 
more partonic operators in the medium~\cite{Qiu:1990xx} \eg, the gluon field strength operator product $F^{+ \nu}(y) F^+_{\nu} (0) $.
To obtain higher twist contributions, higher orders need to be included. A diagram with 
$2n$  gluon insertions may contribute to twist $m \leq 2n$. 
Issues relating to the generalized factorization~\cite{Qiu:1990xy,col89} of such contributions will not be 
dealt with in this effort; the focus will be to obtain an effective description of the 
propagation of a hard parton in transverse color fields. 
In all calculations, the high energy and hence small $g$ limit will be assumed, as a result 
all diagrams containing the four-gluon vertex as in the left panel of Fig.~\ref{four_gluon}. are 
suppressed compared to diagrams with the same number of gluon insertions on the hard line 
which are directly connected to the soft matrix elements. As a result, all such diagrams will be
ignored. 

\begin{figure}[htbp]
%\begin{center}
%  \epsfxsize 80mm
%\hspace{0cm}
  \resizebox{2in}{2in}{\includegraphics[1.5in,0in][5.5in,4in]{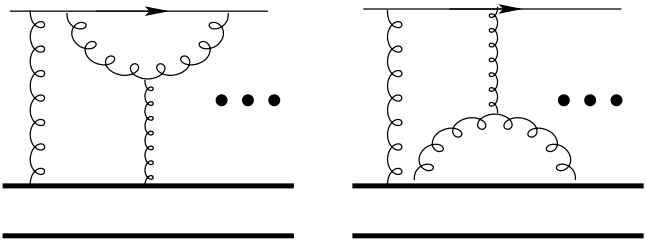}} 
%\vspace{0.25cm}
    \caption{ Left panel shows a virtual contribution to a radiative correction. Right 
panel is a gluon fusion contribution, which is suppressed outside the saturation region.}
    \label{three_gluon}
%  \end{center}
\end{figure}

The aim is to isolate the higher twist contributions at a given order and twist that carry the largest 
multiple of length $L \sim A^{1/3}$.  In the language of power counting, one looks at the combination 
$\A_s^m L^n$ (usually $n \leq m$) and focuses on diagrams with the maximum $n$.
The multiples of $L$ are obtained by insisting on conditions that lead to the largest number of 
propagators going close to their on-shell conditions. This is easily achieved by the hard parton 
having the maximum number of space like exchanges with the medium.  
Such diagrams may or may not include radiated gluons. 
The introduction of radiated gluons leads to the
calculation of energy loss of the parent parton~\cite{HT}. Computation of diagrams without radiated 
gluons deal with the propagation of the parent parton without radiative energy loss. Diagrams where a single 
gluon line originates in the medium and then splits into two prior to attaching with the 
hard parton (left panel of Fig.~\ref{three_gluon}) represent virtual corrections to 
radiative diagrams and fulfill unitarity conservation at 
higher order. While radiative contributions usually require at least one propagator to 
be off-shell, they are enhanced by large logarithms and thus represent a somewhat  
different power counting.
Such diagrams along with their real counterparts will be dealt with in a future publication.. 

Diagrams where multiple gluon lines fuse into a single 
gluon prior to attaching with the hard parton (right panel of Fig.~\ref{three_gluon}) 
do not produce the same length enhancement as 
that of  the individual gluons attaching directly to the hard parton. The reason behind this is that if the 
intermediate gluon line is on-shell then it forces the next quark propagator to go off shell or
vice-versa. In either case, it is suppressed by one factor of $L$ (this is treated in more detail in Appendix A) . 
There is an exception to 
this condition that occurs in the region where the forward momentum fraction of the gluon $x$ is 
very small. In such cases, due to the saturation mechanism, soft gluon populations may be enhanced~\cite{Casalderrey-Solana:2007sw} and 
such diagrams may indeed produce considerable contributions~\cite{Mueller:1989st,McLerran:1993ni}. 
The current analysis will be 
assumed to be carried out outside the saturation regime. As a result, such gluon fusion contributions 
will be ignored.

\begin{figure}[htbp]
%\begin{center}
%  \epsfxsize 80mm
%\hspace{0cm}
\resizebox{3.2in}{2in}{\includegraphics[0in,0in][8in,5in]{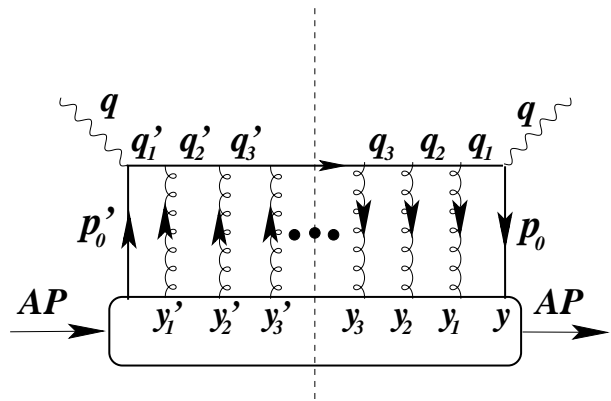}} 
%\vspace{0.25cm}
\caption{ An order $2n$ contribution to $W^{\mu \nu}$. This contributes to twist 
$m \leq 2n$. }
    \label{fig3}
%  \end{center}
\end{figure}

The diagrams 
under discussion have the general form of Fig.~\ref{fig3}. The ellipses in between 
gluon lines in Fig.~\ref{fig3} are meant to indicate an arbitrary number of insertions. Note, there is 
no particular ordering of the gluon lines. Also, the gluon lines are not meant as propagators 
but rather as field insertions at a point $y_i$; hence there is no meaning associated with 
crossed gluon lines. The entire set of $n + n^\p$ vertex insertions (with the gluon vector potentials 
contracted with the nucleus) 
may then be connected by quark propagators in $(n +n^\p)!$ ways 
(this overall combinatoric factor is removed by
 the $(n+n^\p)!$ which appears in the denominator from 
the perturbation expansion).  The reader may question the focus on this sub-class of 
possible diagrams at order $n+n^\p$, where all gluon field operators are contracted with the 
nuclear state. Diagrams of the same order, where the gluon 
field operators are contracted with each other will  contain cut (or uncut) gluon lines, and 
these will represent real (or virtual) contributions to gluon radiation diagrams. A sub class of 
contributions such as those in Fig.~\ref{three_gluon} may indeed be included by a redefinition 
of the effective gluon vector potential. As such, they would represent small additive contributions 
which do not influence the leading length enhanced behavior of the diagrams of Fig.~\ref{fig3} and 
will not be discussed further.

At the $(n + n^\p)$ order, we envisage a general contribution where there are $n$ incoming  gluon 
lines originating in the locations $y_1, \ldots, y_n$ in the amplitude, whereas there are $n^\p$ gluon lines at locations  
$y^\p_1,  \ldots,  y^\p_{n^\p}$ in the complex conjugate. The cut line is 
generically labelled as $l$. The Feynman rule for such a contribution to the hadronic tensor 
is given as, 

\begin{eqnarray}
W^{\mu \nu} &=& \int d^4 y \frac{d^4 l}{(2\pi)^4}  
\prod_{i=1}^n \prod_{j=1}^{n^\p} \left\{ d^4 y_i d^4 y_j^\p 
\frac{d^4 q_{i} d^4 q_j^\p }{(2 \pi)^8} \right\}  \nn \\
\ata   g^{n+n^\p}  \lc A;p | \psi(0) \psibar(y) \g^\mu  
\Pi_{i=1}^n \left[ \frac{\f q_i \g^{\A_i}}{q_i^2  - i \e}   \right] \nn \\
\ata \! \fs l   2\pi \kd(2 l^+ l^- - l_\perp^2)  
\Pi_{j=n^\p}^1 \left[ \frac{\g^{\A_j^\p} \f q_j^\p}{(q_j^{\p})^2  + i \e} \right]
t^{a_i} A_{\A_i}^{a_i}(y_i) \nn \\
\ata t^{a^\p_j} A_{\A_j^\p}^{a^\p_j}(y^\p_j) \g^{\nu} | A;p \rc 
e^{i q \x y } e^{ -i \sum_{i=1}^{n} q_i \x ( y_{i-1}  - y_i )  } \nn \\
\ata e^{-i \sum_{j=1}^{n^\p}   q^\p_j \x ( y^\p_j - y^\p_{j-1}   ) } 
e^{-i l \x (y_n - y^\p_{n^\p})}.  \label{W_mu_nu_general} 
\end{eqnarray}

\nt
The momenta, $q_i, q^\p_j$ label the momenta of the 
various fermion lines. Note, there is overall 
momentum conservation within the diagram. 
As a result, a shift in all momenta may be executed as 
a consequence of momentum conservation at each vertex: 
$q_i = \sum_{m=0}^{i-1} p_m + q$ and 
$q^\p_j = \sum_{k=0}^{j-1} p^\p_{k} + q$, where $p_i,p^\p_j$ 
are the momenta brought in by the 
$i^{\rm th}$ line in the amplitude and the $j^{\rm th}$ line in the 
complex conjugate. It should be pointed out that all such $p_i,p^\p_j$ 
integrals run from $-\infty$ to $\infty$ and thus may indicate 
both incoming and outgoing momenta. Using these substitutions, 
the phase factors in Eq.~\eqref{W_mu_nu_general} may be written 
as

\bea
\Gamma &=& \prod_{i=0}^{n-1}\exp[-i p_i \x y_i  ] \prod_{j=0}^{n^\p -1} 
\exp[ i p^\p_j  \x y^\p_j] \nn \\
\ata \exp\left[ -i y_n \x \left\{ l - \left(q + \sum_{i=0}^{n-1} p_i 
\right) \right\} \right] \nn \\
\ata \exp\left[i y^\p_{n^\p} \x \left\{ l - \left(q + \sum_{j=0}^{n^\p - 1} %\\
p^\p_j \right) \right\} \right], \label{full_phse}
\eea
\nt
where, it is understood that $y^\p_0$ is the origin and $y^0 \equiv y$. 
At this point an $n^{\rm{th}}$ momentum may be introduced, via 

\begin{equation}
1 = \int d^4 p_n  \kd^4 \left( l - \sum_{k=0}^{n} p_k - q \right). \label{pn_delta}
\end{equation}
\nt
This leads to a considerable simplification of the phase factor as 

\begin{eqnarray}
\Gamma &=& \exp\left[ - \sum_{i=0}^n i p_i \x y_i + 
\sum_{j=0}^{n^\p - 1} i p^\p_j \x y^\p_j \right. \nn \\ 
&+& \left. i y^\p_{n^\p} \x \left( \sum_{i=0}^{n} p_i - 
\sum_{j=0}^{n^\p-1} p^\p_j \right) \right]. \label{simple_phse}
\end{eqnarray}
\nt
Note the complete absence of the hard photon momentum $q$ 
from the phase factor.

The approximations stemming from collinear dynamics may 
now be instituted. The calculation is carried out in the 
Breit frame at very high energy. As a result, all momentum 
lines that originate in the target are dominated by the 
large $+$ components of their momentum, followed by their 
transverse coordinates, \tie,

\bea
p_i^+ >> {p_i}_\perp >> p_i^-.
\eea
\nt
In most cases the above condition will allow us to practically 
drop all $-$ components of momentum from the expression for the 
hadronic tensor and focus solely on the $+$ and $\perp$ components.
This procedure will be carried out not only for the momenta but also 
for the field operators. In this effort, calculations will be carried out 
in the covariant gauge. In covariant gauge calculations in the Breit frame at very high 
energy, the dominant components of the vector potential are the forward or 
$(+)$ components~\cite{lqs}, \tie, $A^\sg \sim g^{- \sg} A^{+}$. As there are 
no cancellations involving the vector fields, there will arise no 
further need to consider the $\perp$ components. In the corresponding 
calculation in light-cone gauge $A^{+} = 0$ and the largest components 
are the $A^{\perp}$ components. It should be pointed out that the 
$-$ components are only being dropped from locations where they 
appear in addition to the larger $+,\perp$ components, \tie, they 
are not dropped from the phase factors. 
The collinear approximation may also be used to simplify the quark 
operators, as 

\bea
\lc \psi(0)\psibar(y) \rc  &\simeq&  \frac{1}{4} \left[ 
\g^- \lc \psibar(y) \g^+ \psi(0)\rc  \right. \nn \\
&+& 
\left. \g^{\perp}  \lc \psibar(y) \g_{\perp} \psi(0) \rc \right]
\eea
\nt
where, once again due the high energy limit, all terms collinear with the 
$+$ direction are enhanced by boost compared to the $\perp$ direction and 
thus the second term above maybe dropped. 

The hadronic tensor of Eq.~\eqref{W_mu_nu_general} may be expressed as the 
formal convolution of four terms, 

\bea
W^{\mu \nu} &=& \int d^2l_\perp\md y \md p T(\fp) \Gamma(\fp,\fy) M(\fy)  \nn \\
\ata \kd^2 \left( \vec{l}_\perp - \slm_{i=0}^n \vec{p}^i_\perp \right), \label{general_form}
\eea
\nt
where, $T(\fp)$ denotes the pure momentum component of the 
integrand, $M(\fy)$ is the pure position dependent multi-operator 
matrix element and $\Gamma (\fp,\fy)$ is the phase factor of Eq.~\eqref{simple_phse} that convolutes 
positions and momenta. The bold face quantities $\fp, \fy$ represent 
an array of momenta and positions, 

\[
\fp \equiv [ p_0,\ldots,p_n;p^\p_0,\ldots,p^\p_{n^\p-1};l ] ,
\]

\[
\fy \equiv [y_0,\ldots,y_n;y^\p_1,\ldots,y^\p_{n^\p}].
\]  
  
\nt
The integration measures $\md p$ and $\md y$ denote a 
product of integrals over the different four vectors 
contained in the arrays above. In Eq.~\eqref{general_form}, the integrations over the 
lightcone components of the cut line $l$ have been performed using two of the four 
delta functions introduced in Eq.~\ref{pn_delta}. The remaining two components of the
transverse momentum integration are 

\bea
 \int d^2l_\perp  \kd^2(\vec{l}_\perp - \vec{K}_\perp) 
&=& \int d^2 l_\perp \kd^2 \left( \vec{l}_\perp - \slm_{i=0}^n \vec{p}^i_\perp \right) 
%
% &=& \int d l_\perp^2 \kd^+ \left( l_\perp^2 - |\vec{K}_\perp|^2 \right),
\eea
\nt
where, $\vec{K}_\perp$ is a representative of the sum of the transverse momenta 
brought in by the $n$ gluon insertions.  
The solely momentum dependent terms in the integrand, post 
collinear approximation, may be expressed as,

\bea
T(p) &=& \tr \left[ \gm \g^\mu \frac{ \gm(p_0^+ + q^+) + \gp q^- - \gt \x p_0^\perp}
{2p_0^+ q^- - Q^2 - (p_0^{\perp})^2 - i\e} \right. \ldots  \nn \\
\ata  \gm \frac{ \gm \left( \slm_{i=0}^{n-1} p_i^+  +   q^+\right) + 
\gp q^- - \gt \x \left(\slm_{i=0}^{n-1} p_i^\perp \right)}
{2 \left( \slm_{i=0}^{n-1} p_i^+ \right)q^- 
- Q^2 - \left( \slm_{i=0}^{n-1} p_i^\perp \right)^2 - i\e} \nn \\
\ata \gm \left( \slm_{i=0}^{n} p_i^+  +   q^+ \right)  
+ \gp q^- - \gt \x \left(\slm_{i=0}^{n} p_i^\perp \right) \nn \\
\ata 2\pi \kd\left\{ 2q^- \left(  \slm_{i=0}^n p_i^+ + q^+ \right)  
- \left(\slm_{i=0}^{n} p_i^\perp \right)^2 \right\} \nn \\ 
\ata  \gm \frac{ \gm\left( \slm_{j=0}^{n^\p -1} {p^\p}_j^+  +   q^+\right) + 
\gp q^- - \gt \x \left(\slm_{j=0}^{n^\p-1} {p^\p}_j^\perp \right)}
{2 \left( \slm_{j=0}^{n^\p-1} {p^\p}_i^+ \right)q^- 
- Q^2 - \left( \slm_{i=0}^{n^\p-1} {p^\p}_i^\perp \right)^2 + i\e} \nn \\
\ata \left. \ldots \frac{ \gm({p^\p}_0^+ + q^+) + \gp q^- - \gt \x {p^\p}_0^\perp}
{2{p^\p}_0^+ q^- - Q^2 - ({p^\p}_0^{\perp})^2 + i\e} \g^{\nu }\right],  \label{full_trc}
\eea
\nt
which is the short distance momentum dependent part of the integrand. 
The ellipses in the above equation are meant to indicate the presence 
of propagators with a growing number of additive momentum factors brought in 
by the gluon insertions. 
Note that all appearances of the $(-)$components of the momenta $(p_i^-)$
have been ignored in the above equation. 
The long distance non-perturbative position dependent part of the integrand is given as

\bea
M(y) &\equiv&\tr  \left< A;p \left| \psibar(y) \gp \psi(0) \prod_{i=1}^n t^{a_i} A^+_{a_i} (y_i)  \right. \right. \nn \\
\ata \left. \left. \prod_{j=n^\p}^{1} t^{a^\p_j} {A^\p}^+_{a^\p_j} (y^\p_j) \right| A;p\right> . \label{full_mat}
\eea
\nt
While, the phase factor which convolutes both these terms is essentially contained in Eq.~\eqref{simple_phse}.

The complete absence of  the ($-$) components of the momentum, from all expressions except for the 
phase factors allows for the $p^-$ and ${p^\p}^-$ integrations to be done, resulting in the localization of the 
process on the negative light-cone, \tie, 

\bea
\Gamma^- &=&
\prod_{i=0}^n \prod_{j=0}^{n^\p - 1} \int  d{p^\p}_j^- d p_i^-
e^{-\slm_{i=0}^n ip_i^- (y_i^+ - {y^\p}^+_{n^\p}) } \nn \\
\ata e^{\slm_{j=0}^{n^\p - 1} i{p^\p}_j^- ({y^\p}_j^+ - {y^\p}^+_{n^\p}) } \nn \\
&=& \prod_{i=0}^n \kd( y_i^+ - {y^\p}^+_{n^\p} )
\prod_{j=0}^{n^\p - 1} \kd( {y^\p}_j^+ - {y^\p}^+_{n^\p} ). 
\eea

\nt
In the above equation, ${y^\p}_0 = 0$ and as may be noted from the 
definition of the location arrays is not being integrated over. 
As a result, this 
constrains all the negative light cone locations in the 
above equation to the origin.

The terms in the numerators  of  Eq.~\eqref{full_trc} may be simplified with observation that 
$\gm \gm = \hf g^{- -} = 0$, as a result only the $\gp$ and $\gt$ terms survive the spin sum. 
In the extremely 
high energy limit where $q^- \gg p_i^\perp {p^\p}^\perp_j$, all terms proportional to the 
transverse components of the $\Gamma$ matrices may also be ignored. The terms in the 
denominators may be simplified further with the replacements 
\bea
Q^2 = 2x_B p^+ q^-\,; \,p_i^+ = x_i p^+  \,;\, {p^\p}^+_j = x^\p_j p^+ \label{long_def}  \\
\slm_{k=0}^{i} 2p_\perp^i \cdot p_\perp^k + |p_\perp^i|^2 = 2x^i_D p^+q^-;  \label{i_perp_def} \\
\slm_{l=0}^{j} 2 {p^\p}_\perp^j \cdot {p^\p}_\perp^l + |{p^\p}_\perp^j|^2 = 2 {x^\p}^j_D p^+q^- . \label{j_perp_def}
\eea
\nt
In the above equation, $i$, the index of the unprimed momenta 
(both for the longitudinal and transverse components) runs from $0$ to $n$, 
whereas, $j$, the index of the primed momenta runs from $0$ to $n^\p - 1$, \tie, 
one less than maximum. 

With the above simplifications, the momentum dependent part 
of the hadronic tensor $W^{\mu \nu}$, as sketched in Eq.~\eqref{full_trc}
assumes the form, 

\bea
T(p) &=& 4 \frac{ g^{\mu -} g^{\nu +} + g^{\mu +} g^{\nu -} - g^{\mu \nu}}{(2p^+)^{n+n^\p+1}}  2^{(n+n^\p)}\nn \\
\ata  \left[ x_0 - x_B - x_D^0 - i\e \right]^{-1}   \ldots \nn \\
\ata \!\!\!\!  \left[\slm_{i=0}^{n-1} (x_i - x_D^i) - x_B - i\e \right]^{-1}  \nn \\
\ata 2\pi \kd \left\{ \slm_{i=0}^{n} (x_i - x_D^i) - x_B \right\} \nn \\
\ata \!\!\!\! \left[\slm_{j=0}^{n^\p - 1} (x^\p_j - {x^\p}_D^j) - x_B + i\e \right]^{-1} \ldots \nn \\
\ata \left[ x^\p_0 - x_B - {x^\p}_D^0 + i\e \right]^{-1}. \label{simple_T}
\eea

The integrals over the momenta $p_i^+, {p^\p}_j^+$ may be re-expressed in terms of 
momentum fractions, \tie, 
$ d p_i^+= p^+  dx_i  $.
The $n+n^\p + 1$ integrals over ($+$) momenta in $\md p$ lead to 
an overall factor of ${(p^+)}^{n+n^\p+1}$, which is cancelled by the similar factor 
appearing in the denominator of Eq.~\eqref{simple_T}. As a result, the integral 
measure now has the appearance, 

\bea
\md y \md p  &\equiv& \prod_{i=0}^{n} dy_i^- d^2 {y_i}_\perp 
\prod_{j=1}^{n^\p } d {y^\p}_j^- d^2 {y^\p_j}_\perp  \nn \\
\ata \prod_{i=0}^{n} \frac{dx_i}{2\pi} \frac{d^2 {p_i}_\perp }{(2\pi)^2} 
\prod_{j=0}^{n^\p-1} \frac{dx^\p_j}{2\pi} \frac{d^2 {p^\p_j}_\perp }{(2\pi)^2} .
\eea

The remnant phase factor may be decomposed into a longitudinal piece and a transverse piece, 
which depends on the corresponding components of the momentum appearing as arguments, 
\tie,

\bea
\Gamma &=& \Gamma^+ \Gamma_\perp \equiv \prod_{i=0}^n e^{-i x_i p^+ (y_i^-  - {y^\p}_n^- )} 
\prod_{j=0}^{n^\p - 1} e^{ i x^\p_j p^+ ( {y^\p }_j^- - {y^\p}_n^- )  } \nn \\
\ata \prod_{i=0}^n e^{ i {p_i}^\perp  \x  (y_i^\perp  - {y^\p}_n^\perp )}
\prod_{j=0}^{n^\p-1} e^{ - i  {p_j^\p}^\perp  \x ( {y^\p }_j^\perp - {y^\p}_n^\perp )  }
\eea

The cut  line, in the diagram 
of Fig.~\ref{fig3} is indicated by the delta function appearing in Eq.~\eqref{simple_T}. 
Integrating over the last momentum fraction $x_n$ with the aid of the 
delta function leads to the condition that 

\bea
x_n = x_B + \slm_{i=0}^{n} x_D^i - \slm_{i=0}^{n-1} x_{i}. \label{eql_1}
\eea
\nt
Instituting this condition leads to the separation of the longitudinal phase factor $\Gamma^+$ 
into a left and right piece, \tie, 

\bea
\Gamma^+ &=& \exp \left[ -i\left( x_B + \slm_{i=0}^n x_D^i  \right) p^+ (y_n^-  - {y^\p}_n^-  )\right] \nn \\
\ata \prod_{i=0}^{n-1} \exp \left[ -i  x_i p^+ (y_i^-  - y_n^-  )\right] \nn \\
\ata \prod_{j=0}^{n^\p-1} \exp \left[ i  {x^\p}_j   p^+ ({y^\p}_j^-  - {y^\p}_{n^\p}^-  )\right],
\eea
\nt
where, the second line in the equation above involves only momentum fractions and 
locations from the left-hand side of the cut, where as the last line involves momentum 
fractions and locations from the right hand side of the cut line. 
The integrations over the remaining longitudinal momentum fractions, may now be performed 
starting from the propagators adjacent to the cut and proceeding to the propagators adjacent to 
the photon vertices.  

The first such integration, involves the propagator from the 2nd line of Eq.~\eqref{simple_T}. 
Isolating the piece that depends on the fraction $x_{n-1}$ yields the integral, which may 
be performed by closing the contour of $x_{n-1}^-$ with a counterclockwise semi-circle in 
the positive imaginary direction,  

\bea
&& \int \frac{dx_{n-1}}{2\pi} \frac{e^{-ix_{n-1} p^+ (y_{n-1}^- - y_n^- ) }} 
{ x_{n-1} + \slm_{i=0}^{n-2} (x_i  - x^i_D)  - x^{n-1}_D  - x_B - i\e } \nn \\
&=&  i  \h(  y_n^- - y_{n-1}^- )  \nn \\
\ata e^{  -i \left[  x^{n-1}_D  + x_B - \slm_{i=0}^{n-2} (x_i  - x^i_D)   \right]    
p^+ (y_{n-1}^- - y_n^- ) }.
\eea
\nt
The effect of performing the above integration is the incorporation of 
both the real and imaginary parts of the above propagator into the 
overall expression of the hadronic tensor. It has the physical effect of 
propagating the quark from $y_{n-1}^-$ to $y_n^-$. 
Similarly, the integration over the propagator to the immediate right of the cut line 
may be carried out by closing the contour of $x^\p_{n^\p-1}$ with a clockwise semi-circle in 
the negative imaginary direction: 

\bea
&& \int \frac{dx^\p_{n^\p-1}}{2\pi} \frac{e^{ix^\p_{n^\p-1} p^+ ({y^\p}_{n^\p-1}^- - {y^\p}_{n^\p}^- ) }} 
{ x^\p_{n^\p-1} + \slm_{j=0}^{n^\p - 2} (x^\p_i  - {x^\p}^i_D)  - {x^\p}^{n^\p-1}_D  - x_B + i\e } \nn \\
&=& -  i  \h(  {y^\p}_{n^\p}^-  - {y^\p}_{n^\p-1}^- )  \nn \\
\ata e^{  i \left[  {x^\p}^{n^\p-1}_D  + x_B - \slm_{j=0}^{n^\p-2} ( x^\p_i  - {x^\p}^i_D )    \right]    
p^+ (  {y^\p}_{n^\p-1}^-    -     {y^\p}_{n^\p}^- ) }.
\eea
\nt
Incorporation of the results of the above two  integrals into the longitudinal 
phase factors leads to the expression, 

\bea
\Gamma^+ &=& \exp\left[     -i x_D^{n} p^+ y_n^-  + i {x^\p}_D^{n^\p} p^+ {y^\p}^-_n    \right. \nn \\
&+&  -ix_B p^+ y_{n-1}^- +  ix_B p^+ {y^\p}_{n^\p - 1}^-  \nn \\
&+&  \left. 
-i \slm_{i =0}^{n-1} x^i_D p^+ y_{n-1}^- + i \slm_{j=0}^{n^\p - 1} {x^\p}^j_D p^+ {y^\p}_{n^\p-1}^-  
\right] \nn \\
\ata \prod_{i=0}^{n-2} \exp \left[ -i  x_i p^+ (y_i^-  - y_{n-1}^-  )\right] \nn \\
\ata \prod_{j=0}^{n^\p-2} \exp \left[ i  {x^\p}_j   p^+ ({y^\p}_j^-  - {y^\p}_{n^\p-1}^-  )\right],
\eea
\nt
where, overall momentum conservation was invoked to define the new variable 
\bea
{x^\p}^{n^\p}_ D = \slm_{i=0}^{n} x^i_D  -  \slm_{j=0}^{n^\p - 1 } {x^\p}^j_D.
\eea
\nt
In the above expression, $x_D^i$ for $0 < i < n-1 $ are defined in 
Eq.~\eqref{i_perp_def}, and ${x^\p}^j_D$ for $0<j< n^\p -1$ are defined in 
Eq.~\eqref{j_perp_def}. The $n^{\rm{th}}$ transverse fraction $x_D^n$ is set by 
the delta function arising from the cut line and is given as in Eq.~\eqref{eql_1}.

The trend of the longitudinal momentum fraction integrals is now clear and the general 
result for the integrations over the remnant momentum fractions in the phase factor 
may be carried out.   
The full hadronic tensor of Eq.~\eqref{W_mu_nu_general}, post  
integration over all longitudinal fractions, may now be expressed as 

\bea 
W^{\mu \nu } &=& g^{n + n^\p} \int \frac{d^2l_\perp }{(2\pi)^2}  \prod_{i=0}^n d y_i^- d^2 y_\perp^i 
\prod_{j=1}^{n^\p} d {y^\p}_j^- d^2 {y^\p}^j_\perp \nn \\
& & \int \prod_{i=0}^n \frac{ d^2 p^i_\perp}{(2\pi)^2} 
\prod_{j=0}^{n^\p - 1} \frac{d^2 {p^\p}^j_\perp} { (2\pi)^2} 
(2\pi)^2 \kd^2 ( \vec{l}_\perp - \vec{K}_\perp ) \nn \\
\ata \hf \left( g^{\mu - } g^{\nu + } + g^{\mu +} g^{\nu -} - g^{\mu \nu}  \right) \nn \\
\ata  e^{-ix_B p^+ y^-} \prod_{i=0}^n e^{-ix_D^i p^+ y_i^- } 
e^{i p^i_\perp \x  y^i_\perp  } \nn \\ 
\ata \prod_{j=0}^{n^\p} e^{i {x^\p}_D^j p^+ {y^\p}_j^- } 
e^{-i{p^\p}^j_\perp \x  {y^\p}^j_\perp  } \nn \\
\ata \prod_{i=n}^1 \h ( y_i^-   - y_{i-1}^- ) 
\prod_{j = n^\p}^1 \h ( {y^\p}_j^-  -  {y^\p}_{j-1}^- ) \nn \\
\ata \lc A; p | \psibar(y^-,y_\perp) \g^+ \psi(0) 
\tr \left[ \prod_{i=1}^{n} t^{a_i}  A_{a_i}^+ (y_i^-,y^i_\perp) \right. \nn \\
\ata \left. \prod_{j=n^\p}^{1} t^{a_j} A_{a_j}^+ ( {y^\p}_j^-, {y_\p}^j_\perp ) \right]  | A;p \rc.
\label{W_mu_nu_simple}
\eea
\nt
The expression derived above is completely general, in the sense that no assumption 
regarding the nature of the nuclear state has been made. In the next section, a factorization 
of the above hadronic tensor into a part that is solely dependent on hard momenta and 
a part dependent on soft momenta will be carried out and simplifying assumptions regarding 
the nuclear state made.

%%%%%%%%%%%%%%%%%%%%%%%%%%%%%%%%%%%%%%%%%%%%%%%%%%%%%%%%%%%%%%%%%%%%%%%
%%%%%%%%%%%%%%%%%%%%%%%%%%%%%%%%%%%%%%%%%%%%%%%%%%%%%%%%%%%%%%%%%%%%%%%
%%%%%%%%%%%%%%%%%%%%%%%%%%%%%%%%%%%%%%%%%%%%%%%%%%%%%%%%%%%%%%%%%%%%%%%
%%%%%%%%%%%%%%%%%%%%%%%%%%%%%%%%%%%%%%%%%%%%%%%%%%%%%%%%%%%%%%%%%%%%%%%

\section{Factorization and gradient expansion}

%%%%%%%%%%%%%%%%%%%%%%%%%%%%%%%%%%%%%%%%%%%%%%%%%%%%%%%%%%%%%%%%%%%%%%%
%%%%%%%%%%%%%%%%%%%%%%%%%%%%%%%%%%%%%%%%%%%%%%%%%%%%%%%%%%%%%%%%%%%%%%%
%%%%%%%%%%%%%%%%%%%%%%%%%%%%%%%%%%%%%%%%%%%%%%%%%%%%%%%%%%%%%%%%%%%%%%%
%%%%%%%%%%%%%%%%%%%%%%%%%%%%%%%%%%%%%%%%%%%%%%%%%%%%%%%%%%%%%%%%%%%%%%%

Up to this point, no approximation regarding the nature of the state $| A; p\rc$
or the action of the quark and gluon operators on this state has been made. 
We now approximate the  nucleus as a weakly interacting homogeneous gas of nucleons. 
Such an approximation is only sensible at very high energy, where, due to 
time dilation, the nucleons appear to travel in straight lines almost independent of each other 
over the interval of the interaction of the hard probe. In a sense, all forms of correlators  
between nucleons (spin, momentum, etc.) are assumed to be rather suppressed. 
As a result, the expectation of 
the $n+n^\p+2$ operators in the nuclear state may be decomposed as 

\bea
&& \lc A;p | \psibar(y^-,y_\perp)\g^+ \psi(0) \prod_{i=1}^{n+n^\p} A^{+}_{a_i}(y_i)| \A; p\rc \nn \\
&=& A C^A_{p_1}  \lc p_1 | \psibar(y^-,y_\perp)\g^+ \psi(0) \prod_{i=1}^{n+n^\p} A^{+}_{a_i}(y_i)  | p_1 \rc
\nn \\
&+& C^A_{p_1,p_2}  \lc p_1| \psibar(y^-,y_\perp)\g^+ \psi(0) | p_1 \rc \nn \\
\ata \lc p_2 | \prod_{i=1}^{n+n^\p} A^{+}_{a_i}(y_i)  | p_2 \rc + \ldots ,
\eea
\nt 
where, the factor $C^A_{p_1}$ represents the probability to find a nucleon in the vicinity of the location $\vec{y}$, which is a 
number of order unity (it is the probability to find one of  $A$ nucleons distributed in a volume of size $c A$ within a nucleon 
size sphere centered at $\vec{y}$). The remaining coefficients $C^A_{p_1,\ldots}$ represent the weak 
position correlations between different nucleons. The overall factor of $A$ arises from the determination of the origin (the location $0$ in 
the equation above) in the nucleus, 
which may be situated on any of the $A$ nucleons.

It is clear from the 
above decomposition that the largest contribution arises from the term where the expectation of 
each partonic operator is evaluated in separate nucleon states as the $\vec{y}_i$ integrations 
may be carried out over the nuclear volume. 
As a nucleon is a color singlet, any combination of quark or gluon field strength insertions in 
a nucleon state  must itself  be restricted to a  color singlet combination. As a result, the expectation of 
single partonic operators in nucleon states is vanishing. The first (and hence largest) non-zero contribution 
emanates from the terms where the quark operators in the singlet color combination are evaluated in a 
nucleon state and the $n+n^\p$ gluons are divided into pairs of singlet combinations, with each singlet 
pair evaluated in a separate nucleon state. This requires that $n+n^\p$ is even and may lead to 
a maximum overall factor of 

\bea
C^A_{p_1,p_2,\ldots}  \sim  A^{[(n+n^\p+2)/2] } ,
\eea
\nt
in the large $A$ limit.  It should be pointed out that large contributions may also arise, in principle, 
when $n+n^\p$ is odd. In this case, the two quarks and a gluon are considered in the singlet 
combination with the remaining 
gluons evaluated in singlet pairs in the remaining nucleons. Here we institute the experimental 
observation that $(n)$-parton observables are much smaller than $(n-1)$-parton observables. 
This is only true, once again, outside the saturation regime, as discussed at the beginning of this 
section. Such odd parton expectations also become important in cases where nucleon polarization 
effects are being studied where the spins of the nucleons are not averaged over~\cite{Ji:2006br}. 
In this effort, the focus remains exclusively outside such regions, as a result we ignore all terms 
with more than two quarks or two gluons per nucleon.

Further simplifications arise in the evaluation of  gluon pairs in a singlet combination in the nucleon 
states by carrying out the $y_\perp$ integrations. The basic object under consideration is  (ignoring the 
longitudinal positions and color indices on the vector potentials)

\bea
\mbx & & \int  d^2 y^i_\perp  d^2 {y^\p}^j_\perp \lc p | A^{+}  (\vec{y}^i_\perp )  A^{+}  ( \vec{y^\p}^j_\perp ) | p \rc \nn \\
&\times & e^{-i x_D^i p^+ y_i^-}  e^{ i p^i_\perp \x y^i_\perp}   
e^{i {x^\p}_D^j p^+ {y^\p}_j^-}  e^{ - i {p^\p}^j_\perp \x {y^\p}^j_\perp}  \nn \\
&=& (2\pi)^2 \kd^2( {\vp}^{\,i}_\perp - {\vec{p^\p}} ^j_\perp )  \int d^2 y_\perp  e^{-i x_D^i p^+ ( y_i^-  -  {y^\p}_j^- )} \nn \\
% \frac{-g^{\A \B}_{\perp}}{2} \nn \\ 
%
&\times& 
 e^{ i p_\perp \x y_\perp} \lc p | A^{+} (\vec{y}_\perp/2 )  A^{+}  ( - \vec{y}_\perp/2 ) | p \rc , \label{two_gluon_cor}
 \eea 
where, $y_\perp$ is the transverse gap between the two gluon insertions and $p_\perp = (p^i_\perp + p^j_\perp)/2$. 
The physics of the 
above equation is essentially the transverse translation symmetry of the two gluon correlator 
in a very large nucleus. One will note that the two dimensional delta function over the transverse 
momenta has removed an integration over the transverse area of the nucleus thus reducing the 
overall $A$ enhancement that may be obtained. This is then used to equate the transverse 
momenta emanating from the two gluon insertions in the amplitude and complex conjugate amplitude. 
This also simplifies the longitudinal phase factors which now depends solely on the difference of the 
longitudinal positions of the two gluon insertions.

Up to this point, the collinear approximation has been used to simplify the 
expressions for the hadronic tensor, without the introduction of factorization. 
The separation of the hadronic tensor into a hard short distance piece and a soft 
long distance contribution may now be accomplished. All factors in Eq.~\eqref{W_mu_nu_simple} 
which contain the hard scales $p^+,q^-$ constitute the hard part. All factors that
depend solely on the soft $\perp$ momenta and distances along with the matrix element 
constitute the long distance element, \tie, the first part of the integrand in the $3^{\rm{rd}}$ and 
$4^{\rm{th}}$  lines in Eq.~\eqref{W_mu_nu_simple} belongs in the hard part along with all phase 
factors which contain a factor $ x_D^i p^+ $ or $ {x^\p}_D^j p^+$ as part of their arguments. 
The purely transverse phase factors such as $\exp [ i \vec{p}_\perp \x \vec{y}_\perp] $  belong in the 
soft part along with the matrix elements. Thus, we may decompose the hadronic tensor as, 

\bea 
W^{\mu \nu} = \int \md y \md p_\perp H(p^+,q^-,p_\perp,y)  S(p_\perp, y) .
\eea
\nt
In the above equation, $y$ and $p_\perp$ are representative of the entire set of distances and transverse 
momentum that appear in Eq.~\eqref{W_mu_nu_simple}. One now invokes the collinear 
approximation in expanding the hard part as a Taylor expansion in transverse 
momenta around the origin $p_\perp^i \ra 0$. In the case of a symmetric cut with $n=n^\p$ there are 
$2n$ gluon insertions and as a result, as many derivatives, which involve $n$ different transverse momenta.
All terms of the form 
\[
 \prod_{i=1}^m \frac{1}{2} \frac{\prt^2  }{ \prt {p_\perp^i}^\A \prt {p_\perp^i}^\B} 
\left.  H \right|_{p_\perp = 0}{p_\perp^i}^\A {p_\perp^i}^\B  ,
\]
where, $m<n$, yield gauge corrections for the contributions with $2m$ 
gluon insertions and as many derivatives~\cite{lqs}. The genuine $2n$ twist correction at this 
order has $m=n$.  Expanding to this order, one obtains the 
generic term,

\bea
& & \prod_{i=1}^n \frac{1}{2} \frac{\prt^2  }{ \prt {p_\perp^i}^\A \prt {p_\perp^i}^\B} 
\left. H(p^+, q^-, p_\perp^i, )\right|_{p_\perp^i = 0 } \nn \\
\ata {p_\perp^i}^\A {p_\perp^i}^\B   A^+_{a_i} (y_i^-,y_\perp^i/2)
 A^+_{a^\p_j} (  {y^\p}_j^-   ,   -y_\perp^i/2   ). \label{2n_der}
\eea
\nt
Where, we have assumed the result of Eq.~\eqref{two_gluon_cor} and reintroduced the 
color indices and longitudinal locations.

Using integration by parts over the transverse distance $y_\perp^i$ one may convert 
the product ${p_\perp^i}^\A A^+_{a_i} (y_i^-,y_\perp^i/2) \ra \frac{1}{2} \prt_\perp^\A A^+_{a_i} (y_i^-,y_\perp^i/2)$ .
In the extreme collinear limit, in the presence of a hard scale such that $g$ is small, one may make the approximation,

\bea 
\prt_\perp A^+_a \simeq {F_a}^+_\perp, \label{glue_field}
\eea
where, ${F_a}^+_\perp$ represents the gluon field strength.  Carrying this out consistently on the 
two gluon operator in the nucleon state of Eq.~\eqref{two_gluon_cor} and ignoring derivatives of 
the field strength [$\prt^\A F^{+ \B} \sim g(m^2) g^{\A \B} j^+ \ra 0$] we obtain, 
\bea
\mbx \!\!\!\!\!\!\!\!\!\!\!\! && 
\int d^2 y_\perp {p_\perp^i}^\A {p_\perp^i}^\B  e^{ i p_\perp \x y_\perp} \lc p | A^{+} (\vec{y}_\perp/2 )  A^{+}  ( - \vec{y}_\perp/2 ) | p \rc  \nn \\
\mbx \!\!\!\!\!\!\!\!\!\!\!\! &=& 
\int d^2 y_\perp  e^{ i p_\perp \x y_\perp} \frac{1}{2} \lc p | F^{+ \A} (\vec{y}_\perp/2 )  F^{+ \B}  ( - \vec{y}_\perp/2 ) | p \rc \nn \\
\mbx \!\!\!\!\!\!\!\!\!\!\!\!&=& 
\int d^2 y_\perp  e^{ i p_\perp \x y_\perp} \frac{-g_\perp^{\A\B}}{4} \lc p | F^{+ \rho} (\vec{y}_\perp/2 )  F^{+}_{\rho}  ( - \vec{y}_\perp/2 ) | p \rc.
 \eea 

\nt
In the last line of the above equation we have averaged over the spins in the two field strength expection in the nucleon state with the constraint that 
the operator being evaluated in the nucleon be a spin singlet. The nucleon states are always assumed to be spin singlets or in spin averaged states.

The hard part which consists of $2n$ transverse momentum derivatives may now be simplified further
\bea
\prod_i  \frac{\prt^2}{\prt {p^i_\perp}^2} \kd^2 (\vec{l}_\perp - \vec{K}_\perp) 
=  \left( \nabla_{l_\perp}^2 \right)^n \kd^2 (\vec{l}_\perp - \vec{K}_\perp)  . \label{n_equal_nprime}
\eea
With the derivative 
expansion (at vanishing transverse momenta $p^i_\perp \ra 0$) imposed on the hard part $H$, it no longer has any functional 
dependence on the transverse momenta. The integrations over the transverse 
momenta may now be included completely into the soft part. 
The action of the derivatives remains on the transverse $\kd$-function as 
the only non-vanishing contribution from the collinear expansion on  Eq.~\eqref{W_mu_nu_simple}.

In the interest of simplicity, we have considered the symmetric case of gluon insertions above, where $n=n^\p$. 
The other cases may be easily computed in similar fashion (the case where $n=n^\p - 2$ is considered in Appendix B). 
The longitudinal integrals, due to color 
confinement  yield the requirement that the longitudinal locations of the two gluons which act on the same 
nucleon state be in close 
proximity. One now tries to identify the most length enhanced term by isolating the maximum number of 
unconstrained $dy^-$ integrals.  Note that, due to color confinement (ignoring color and spin indices), 

\bea
&& \int d  y^- d {y^\p}^- d^2 y_\perp \frac{d^2 p_\perp}{(2\pi)^2}   e^{-i x_D^i p^+ ( y^-  -  {y^\p}^- )} \\
% \frac{-g^{\A \B}_{\perp}}{2} \nn \\ 
%
&\times& 
 e^{ i p_\perp \x y_\perp} \lc p | F  (y^- , y_\perp/2)  F ({y^\p}^- , -y_\perp/2) | p \rc \nn \\
&=&  \int d Y^-  \int d \kd y^- d^2 y_\perp \frac{d^2 p_\perp}{(2\pi)^2}   e^{-i x_D p^+ \kd y^- } \nn \\ 
&\times& e^{ i p_\perp \x y_\perp}  \lc p | F  (\kd y^-/2 , y_\perp/2)  F ( -\kd y^-/2 , -y_\perp/2) | p \rc  \nn
\eea
where, $x_D = \frac{p_\perp^2}{2 p^+ q^-}$ is a function of the $p_\perp$ itself.

In the equation above, all three integrals over $ \kd y^- , \vec{y}_\perp$ are limited by the nucleon size. However, the integral 
over $Y^-$, under the assumption of longitudinal translation invariance of the two point correlator, may span 
the entire length of the nucleus. 
Each such integral yields a factor of $L^- \sim A^{1/3}$ from the unconstrained $Y^-$ integration.  
The equating of the pairs of transverse momenta 
that appear in each two-gluon correlation, as 
well as the relation between the longitudinal momenta from the $\h$-functions in Eq.~\eqref{W_mu_nu_simple}, 
require that the largest transverse momentum broadening and largest length enhancement arises from the terms 
where the gluon correlations are built up in a mirror symmetric fashion, \tie, where the gluon insertion at $y^i$ is 
contracted with that at ${y^\p}^i$.  This reduces the focus on terms where $n=n^\p$ which produce a transverse 
momentum broadening of order $A^{n/3}$ and provides an \emph{aposteriori} justification for our discussion of symmetric 
terms.  
% One may now let the transverse momenta $p^i_\perp$ (for all $i$) in the hard part 
% tend to zero and integrate over the remaining $p^i_\perp$ integrals in the soft part, setting the corresponding transverse 
% distances between the gluon insertions in a nucleon to zero \tie, $y^i_\perp \ra 0$.

One may average colors of the quark and gluon field operators, 

\bea 
\lc p | F^a F^b | p \rc = \frac{\kd^{ab}}{ (N_c^2 - 1)} \lc p | F^a F^a | p \rc \label{glue_color_conf}\\ 
\lc p | \psibar_i\g^+ \psi_j | p \rc = \frac{\kd_{ij}}{  N_c } \lc p |  \psibar \g^+ \psi| p \rc. \label{quark_color_conf}
\eea 
\nt
This reduces the overall trace over color factors to 
\bea
&& \frac{1}{N_c (N_c^2 - 1)^n} \tr \left[ \prod_{i=1}^n t^{a_i} \prod_{j=n}^1 t^{a_j} \right] \nn \\
&=& \frac{C_F^n}{(N_c^2 - 1)^n}
= \frac{1}{(2N_c)^n}. 
\eea
\nt
If the original parton was a gluon, one would simply replace $C_F$ with $C_A$ in the above 
equation.

The remaining $n$ longitudinal position integrals for the gluon insertions may be simplified as 
\bea
\int \prod_{i=1}^{n} dy_i^-  \h(y_i^- - y_{i-1}^-) = \frac{1}{n!} \int \prod_{i=1}^n dy_i^- .
\eea
\nt
Invoking the above simplifications, the leading length enhanced contribution at order $2n$ to the differential hadronic 
tensor is obtained as (in the following we have replaced $\kd y^-$ with $y^-$ and translations made to simplify the resulting expression),

\bea 
\mbx\!\!\!\!\!\!\!\!\frac{d^2 W_n^{\mu \nu}}{d^2 l_\perp} \!\!\!\! 
&=&  \!\!\!\!C^A_{p_0,\ldots, p_n}W_0^{\mu \nu} \frac{1}{n!} \left[ \{\nabla^2_{l_\perp}\}^n \kd^2 (\vec{l}_\perp) \right] \nn \\
\mbx\!\!\!\!\ata \!\!\!\!\left[ \frac{\pi^2 \A_s  }{2N_c} L^- \int  \frac{d y^- d^2 y_\perp d^2 p_\perp}{(2\pi)^3}   e^{-i x_D p^+  y^- } \right. \nn \\ 
\mbx\!\!\!\!&\times& \left.   \!\!\!\! e^{i p_\perp \x y_\perp}
\lc p | {F^a}^{+ \A} ( y^- , y_\perp)  {F^a}_{\A,}^{\,\,\, +}( 0 ) | p \rc \frac{\mbx^{\mbx}}{\mbx}\right]^n  \label{W_n}
\eea
% \int \frac{dy^-}{2\pi} \lc p | {F^a}^{+ \A} {F^a}_{\A,}^{\,\,\, +}| p \rc \right]^n.  \label{W_n}
% \eea
\nt
In the next section, a re-summation of all such contributions of arbitrary order $n$ will be carried out to calculate the 
leading differential distribution of  a hard jet in transverse momentum as a function of the length traversed in the medium.

%%%%%%%%%%%%%%%%%%%%%%%%%%%%%%%%%%%%%%%%%%%%%%%%%%%%%%%%%%
%%%%%%%%%%%%%%%%%%%%%%%%%%%%%%%%%%%%%%%%%%%%%%%%%%%%%%%%%%
%%%%%%%%%%%%%%%%%%%%%%%%%%%%%%%%%%%%%%%%%%%%%%%%%%%%%%%%%%
%%%%%%%%%%%%%%%%%%%%%%%%%%%%%%%%%%%%%%%%%%%%%%%%%%%%%%%%%%
%%%%%%%%%%%%%%%%%%%%%%%%%%%%%%%%%%%%%%%%%%%%%%%%%%%%%%%%%%

 \section{Re-summation and transverse momentum diffusion}

%%%%%%%%%%%%%%%%%%%%%%%%%%%%%%%%%%%%%%%%%%%%%%%%%%%%%%%%%%
%%%%%%%%%%%%%%%%%%%%%%%%%%%%%%%%%%%%%%%%%%%%%%%%%%%%%%%%%%
%%%%%%%%%%%%%%%%%%%%%%%%%%%%%%%%%%%%%%%%%%%%%%%%%%%%%%%%%%
%%%%%%%%%%%%%%%%%%%%%%%%%%%%%%%%%%%%%%%%%%%%%%%%%%%%%%%%%%

In the preceding section, the leading length enhanced twist-$2n$ contribution was evaluated. 
As is obvious from Eq.~\eqref{W_n}, the essential combination of factors is that included in 
the square brackets. When this factor is not very small, all such terms (for all $n$) need to be re-summed  
into the generalized differential hadronic tensor, which will include contributions at all-twist.

To simplify the re-summation, one further approximation is required. In the preceding section, 
a model of the nucleus as a weakly interacting homogeneous gas of nucleons was used. The 
formal expression  of this assumption is hidden within the dimensionful parameter $C^A_{p_0,\ldots,p_n}$. 
The precise evaluation of such combinatoric coefficients is rather complicated, even for the case of 
next-to-leading twist~\cite{maj04e,Osborne:2002st}.  From general dimensional arguments, in the case of 
non-interacting nucleons, this factor may be approximated as, 

\bea
C^A_{p_0,\ldots, p_n} \simeq C^A_p  \left( \frac{\rho}{2 p^+} \right)^n,
\eea
\nt
where, $\rho$ is the nucleon density inside the nucleus and $1/2p^+$ originates in the 
normalization of  the nucleon state. The remaining unknown coefficient $C^A_p$ 
is now considered to be independent of the order $n$.
Straightforward substitution the above approximation into Eq.~\eqref{W_n}, leads to 
a simplified form for the $2n^{\rm{th}}$ order contribution to the differential 
hadronic tensor. The full hadronic tensor, which includes contributions from all 
orders, may be expressed as,

\bea 
\frac{d^2 W^{\mu \nu}}{d^2 l_\perp} = \sum_{i=0}^\infty \frac{d^2 W^{\mu \nu}_i}{d^2 l_\perp}. 
\eea
\nt
Note that while $i$ runs over all integers, each term refers to an even number ($2i$) of gluon insertions. 
Using the simplifications mentioned above and Eq.~\eqref{W_n} for the $2n^{\rm{th}}$ order term, we 
obtain the generalized differential hadronic tensor, 

\bea
\!\!\!\!\!\!\!\!
\frac{d^2 W^{\mu \nu }}{d^2 l_\perp} = e^{ (D L^-) \nabla^2_{l_\perp} } 
\frac{d^2 W_0^{\mu \nu }}{d^2 l_\perp},  \label{resummed}
\eea
\nt
where, $d^2W_0^{\mu \nu}/d^2 l_\perp$ is given by Eq.~\eqref{d_W_0}, and the constant 
$D$ is given as, 

\bea
D &=& \frac{\pi^2 \A_s}{2 N_c} \rho \int \frac{d^3 y  d^2 p_\perp}{(2\pi)^3 2 p^+} \lc p | {F^a}^{+ \A}(y) {F^a}_{\A,}^{\,\,\, +}(0)| p \rc. \nn \\
% \label{D}
%
\ata \exp \left\{-i \left( \frac{p_\perp^2}{2q^-}  y^- -  p_\perp \x y_\perp \right) \right\} \nn \\ 
&\simeq&  \frac{\pi^2 \A_s}{4 N_c} \rho x_T G(x_T).
\label{D}
\eea
In the above equation, $x_T = \lc p_\perp^2 \rc/(2p^+ q^-)$,  where $\lc p_\perp^2 \rc$  is the average transverse momentum 
that a nucleon may impart to a hard parton passing through it and $ G(x_T)$ is the longitudinal gluon distribution function of the nucleon at 
$x_T$, where

\bea
G(x) = \int \frac{d y^-}{2\pi} \frac{e^{-ixp^+ y^-}}{xp^+} \lc p | {F^a}^{+ \A}(y^-) {F^a}_{\A,}^{\,\,\, +}(0)| p \rc.
\eea
\nt
The scale of the gluon distribution is essentially the same as that used for the scatterings as well as  for the final fragmentation \tie, $m^2$.

Following the form of  Eq.~\eqref{d_W_0}, the re-summed differential hadronic tensor may be expressed in terms of 
a product of the leading order, leading twist, integrated hadronic tensor and a length dependent 
transverse momentum distribution function $\phi (L^-, \vec{l}_\perp)$, 

\bea
\frac{d^2 W^{\mu \nu }}{d^2 l_\perp} = C_p^A W_0^{\mu \nu}\phi(L^-, \vec{l}_\perp) \label{phi}
\eea 
\nt
While our analysis has focused on the hadronic tensor, it should be pointed out that $W^{\mu \nu}$
may be immediately supplanted with the differential cross section to produce a jet with a transverse 
momentum $\vec{l}_\perp$ using Eq.~\eqref{LO_cross}. 
In a sense, the entire process of  the production and subsequent propagation of  the hard quark through 
the nucleus has been factorized into two parts: $C_p^A W_0^{\mu \nu}$ represents the initial production 
in a hard scattering with a virtual photon and $\phi(L^-, \vec{l}_\perp)$ represents its propagation and 
transverse momentum broadening. The entire length ($L^-$) and transverse momentum ($\vec{l}_\perp$) 
dependence is contained entirely within the factor $[\phi(L^-,\vec{l}_\perp)]$.
Differentiating Eq.~\eqref{resummed} with respect to $L^-$ suggests the obvious relation for the 
differential transverse momentum distribution, 

\bea
\frac{\prt \phi(L^-, \vec{l}_\perp ) }{\prt L^-} = D {\nabla^2_{l_\perp}} \phi (L^-, \vec{l}_\perp ), \label{diffusion_eqn}
\eea
\nt
which is a two dimensional diffusion equation in the vector $\vec{l}_\perp$ with the trace of the diffusion tensor 
given by Eq.~\eqref{D}. The initial condition is discerned from Eq.~\eqref{d_W_0} as, 

\bea
\phi(L^-=0, \vec{l}_\perp) = \kd^2(\vec{l}_\perp) \label{initial_cond}
\eea

The two dimensional diffusion equation, with the above mentioned initial condition,  has the well known solution~\cite{sneddon}, 

\bea 
\phi(L^-,\vec{l}_\perp) = \frac{1}{4 \pi D L^-} \exp \left\{-  \frac{l_\perp^2}{4 D L^-} \right\}. \label{solution}
\eea
\nt
Using the above diffusion equation, all moments of the transverse momentum distribution 
may be calculated. The first non-zero moment is the average squared transverse momentum 
distribution at a length $L^-$, 

\bea
\lc l_\perp^2 \rc_{L^-} = \int d^2 l_\perp l^2_\perp \phi (L^-,\vec{l}_\perp) = 4 D L^-.
\eea
\nt
As a result, the transport coefficient of energy loss defined as the average 
transverse momentum squared gained by the jet per unit length traversed in the 
medium is given as (note: $L^- = 2L$ the actual length traversed),

\bea
\hat{q} = \frac{2 \lc l_\perp^2 \rc_{L^-}}{L^-} = 8 D = \frac{2 \pi^2 \A_s }{N_c} \rho x_T G(x_T,m^2)|_{x \ra 0},  
\eea
\nt
% where, the gluon structure function, defined as 
It should be pointed out that in an earlier work where a diffusion equation was used to 
understand the systematics of transverse momentum broadening of jets~\cite{Majumder:2006wi}, 
the diffusion tensor $\bar{D}$ was defined to be four times that in Eq.~\eqref{D}, such that one obtains, 
the mean square transverse momentum broadening as $\lc l_\perp^2 \rc = \bar{D} L^-$ and 
$\hat{q} = 2 \bar{D}$. 

The final numerical value of the scale of the gluon distribution function deserves some discussion. We have 
computed a class of all twist corrections to what amounts to a leading order (LO) hard scattering process. In 
the numerical implementation of such processes, there exists only one large scale \tie, $Q^2$. Thus, while in principle 
the multiple scattering and final fragmentation occur at a softer scale, in LO calculations one may set $m^2 = Q^2$ 
for both the gluon distribution as well as the final fragmentation. 
If the hard scattering were computed at NLO this may no longer be the case.

%%%%%%%%%%%%%%%%%%%%%%%%%%%%%%%%%%%%%%%%%%%%%%%%%%%%%%%%%%
%%%%%%%%%%%%%%%%%%%%%%%%%%%%%%%%%%%%%%%%%%%%%%%%%%%%%%%%%%
%%%%%%%%%%%%%%%%%%%%%%%%%%%%%%%%%%%%%%%%%%%%%%%%%%%%%%%%%%
%%%%%%%%%%%%%%%%%%%%%%%%%%%%%%%%%%%%%%%%%%%%%%%%%%%%%%%%%%
%%%%%%%%%%%%%%%%%%%%%%%%%%%%%%%%%%%%%%%%%%%%%%%%%%%%%%%%%%

\section{Lorentz-Langevin analysis and classical propagation}

%%%%%%%%%%%%%%%%%%%%%%%%%%%%%%%%%%%%%%%%%%%%%%%%%%%%%%%%%%
%%%%%%%%%%%%%%%%%%%%%%%%%%%%%%%%%%%%%%%%%%%%%%%%%%%%%%%%%%
%%%%%%%%%%%%%%%%%%%%%%%%%%%%%%%%%%%%%%%%%%%%%%%%%%%%%%%%%%
%%%%%%%%%%%%%%%%%%%%%%%%%%%%%%%%%%%%%%%%%%%%%%%%%%%%%%%%%%
%%%%%%%%%%%%%%%%%%%%%%%%%%%%%%%%%%%%%%%%%%%%%%%%%%%%%%%%%%

The field theory calculation presented in the preceding sections 
demonstrates that, in the limit of no radiation, the transverse dynamics 
of hard collinear jets passing through dense matter may be understood 
in terms of a transverse momentum diffusion equation. As such, no 
clear physical picture has emerged from all the 
approximations  carried out.  The situation at this point is not 
dissimilar to that encountered in the hard thermal loop (HTL) effective 
theory of QCD~\cite{Braaten:1989mz,Frenkel:1989br} where hard momenta ($p \sim T$)
were integrated out to find effective propagators for soft modes ($p \sim gT$). 
Identical results were also derived if one assumed the more physical picture 
of the hard modes to be classical particles moving under the influence of a soft 
color field which represented the soft modes~\cite{Blaizot:1993zk}. 
In this section, we demonstrate that an almost identical situation is true for the 
case of higher twist jet broadening.

In the following, an alternative calculation of  the transverse broadening of  an energetic jet as it passes 
through a dense colored medium will be carried out classically. It is not a complete 
surprise that such a derivation yields a similar result as the rigorous quantum mechanical 
analysis performed above. In this case, one computes the propagation of  a colored 
particle as it propagates under the action of  fluctuating color fields which have a 
short color correlation length. This last condition is meant to mimic the effect of  the 
color confinement at nucleon like distances as used in the preceding derivation.  

Imagine, as in the preceding section, that a colored particle moves under the 
influence of a color Lorentz force in the $(-)$ direction. The equation of motion 
for a classical particle is, 

\bea
\frac{d p_\mu }{dt} &=& g Q^a (y) F^a_{ \mu \nu } (y)  v^\nu , 
\eea
where, $t$ is the time in the observers rest frame, $Q^a(y)$ is the color charge 
expectation of the particle at location $y$,  $ { F^a }^{ \mu \nu } (y) $ is the 
color field strength at that location and $v$ is the three velocity written in four 
component form \tie, $v \equiv [1,\vec{v}]$ in time-space components. In light cone 
components the vector $v$ is given as, $v \equiv [(1+v_z)/2, 1-v_z, \vec{v}_\perp]$ .
Note that for a light like particle $v_z^2 + v_\perp^2 = 1$, thus $v$ is akin to a unit vector.
For a particle moving predominantly in the negative $z$ direction as in the case 
of the preceding sections, $v_z < 0$ and $|v_z| >> |v_\perp| $.  Hence, one 
may make the approximation, 

\bea
F_{\mu \nu} v^\nu \simeq 2 F_{\mu}^+.
\eea

As a result, the squared transverse broadening in momentum experienced by the 
light-like particle is given as 

\bea 
| \Delta p_\perp (y)|^2 &=& -g_\perp^{\mu \nu } \Delta p_\mu(y) \Delta p_\nu(y) \\
&=& g^2 \int\limits_{y_0}^y d y_1^- d y_2^- Q^a Q^b { F^a }^{\,\,\, +}_{\A,}(y_1^-) 
{ F^b }^{+ \A} (y_2^-) \nn,
\eea
\nt
where, the substitution $y^- = 2t$ has been made for a light like particle traveling in the negative 
$z$-direction.
Taking the expectation of the product of gluon field strength operators, one may now institute the 
condition that the medium has a short color correlation length, \tie,

\bea
& &\int dy_1^- dy_2^- \lc  {F^a }^{\,\,\, +}_{\A,}(y_1^-) { F^b }^{+ \A} (y_2^-) \rc \nn \\
&\sim& \int  d y_1^-  \lc  {F^a }^{\,\,\, +}_{\A,} { F^b }^{+ \A}  \rc  y_{conf}^-.
\eea
\nt
The average over the color of the gluon field 
correlators may now be carried out (ignoring position coordinates), 

\bea
\lc F^a F^b \rc = \frac{\kd^{ab}}{ ( N_c^2 - 1 ) } \lc F^c F^c \rc.
\eea
\nt
The color charges, $Q^a$ represent the expectation of the color matrices in the 
state of the colored parton. As a result $\kd^{ab} Q^a Q^b = C_F$ for a 
quark parton. Hence, one obtains the mean square transverse momentum after a 
parton travels a distance from $y_0^-$ to $y^-$ as given by, 

\bea
\!\!\!\!\!\!\! | \Delta p_\perp (y)|^2 &=& \frac{ 4\pi^2 \A_s t }{N_c} \\
\ata \int \frac{dy^-}{2\pi} 
\left< { F^a }^{\,\,\, +}_{\A,} \left( \frac{y^-}{2} \right)  { F^a }^{+ \A} \left( \frac{-y^-}{2} \right) \right> .\nn
\eea

Using the same normalization as in the preceding section for turning a nuclear state into a gas of
nucleon states \tie, 

\[
\left< \frac{\mbx}{\mbx} \mathcal{O} \right> \ra \frac{\rho}{2p^+} \lc p | \mathcal{O}| p \rc. 
\]
\nt
one obtains identical expressions for the mean square transverse momentum per unit length 
as obtained from the higher twist analysis. This leads to the contention that the re-summation 
of length enhanced twist corrections on to an outgoing parton line is effectively described as the
 classical propagation of 
hard colored partons in soft fields, as long as the correlation lengths of the soft fields are short.

%%%%%%%%%%%%%%%%%%%%%%%%%%%%%%%%%%%%%%%%%%%%%%%%%%%%%%%%%%
%%%%%%%%%%%%%%%%%%%%%%%%%%%%%%%%%%%%%%%%%%%%%%%%%%%%%%%%%%
%%%%%%%%%%%%%%%%%%%%%%%%%%%%%%%%%%%%%%%%%%%%%%%%%%%%%%%%%%
%%%%%%%%%%%%%%%%%%%%%%%%%%%%%%%%%%%%%%%%%%%%%%%%%%%%%%%%%%
%%%%%%%%%%%%%%%%%%%%%%%%%%%%%%%%%%%%%%%%%%%%%%%%%%%%%%%%%%

\section{Discussions and conclusions}

%%%%%%%%%%%%%%%%%%%%%%%%%%%%%%%%%%%%%%%%%%%%%%%%%%%%%%%%%%
%%%%%%%%%%%%%%%%%%%%%%%%%%%%%%%%%%%%%%%%%%%%%%%%%%%%%%%%%%
%%%%%%%%%%%%%%%%%%%%%%%%%%%%%%%%%%%%%%%%%%%%%%%%%%%%%%%%%%
%%%%%%%%%%%%%%%%%%%%%%%%%%%%%%%%%%%%%%%%%%%%%%%%%%%%%%%%%%
%%%%%%%%%%%%%%%%%%%%%%%%%%%%%%%%%%%%%%%%%%%%%%%%%%%%%%%%%%

The propagation and energy loss of hard partons in dense matter is a topic of much 
current interest. In the current effort, a detailed analysis of the transverse momentum diffusion 
of jets produced in DIS off large nuclei, as they pass through cold nuclear matter 
was carried out. Radiation and radiative 
energy loss were ignored, though, in reality, such partons  will tend to radiate and 
lose a substantial fraction of their energy as they traverse the dense matter. 
Complications arising from energy loss were ignored in an effort to isolate the 
systematics of the propagation of a single parton in a fluctuating color field 
without radiation. Radiation will be included in a subsequent effort to complete the 
formulation of  the propagation of hard partons through extended dense matter.

The leading length enhanced higher twist corrections which produce maximal broadening 
were identified and re-summed onto the out going parton line. The final transverse 
momentum distribution at a given length traversed was shown to obey a two-dimensional 
diffusion equation. The diffusion equation was solved and the mean square transverse 
momentum per unit length ($\hat{q}$) calculated. This is the first effort, to the knowledge of the 
authors, which presents a first principles derivation of the transverse momentum distribution 
function (within the higher twist formalism) and a relation between the diffusion tensor 
and the transport coefficient $\hat{q}$. 
Our results differ from those of an earlier attempt~\cite{Fries:2002mu} to re-sum length 
enhanced multiple scattering diagrams, which found a shift, rather than a broadening, of the transverse 
momentum distribution. We believe that our result is more broadly applicable.

The results for $\hat{q}$ obtained from the higher 
twist analysis were shown to be identical to that obtained for a point like colored particle 
moving in a fluctuating color field if the color correlation length of the fields is assumed to 
be short. This places the effective picture of higher twist re-summation on the same footing 
as that of the HTL effective theory which also admits an almost similar picture. This 
suggests the possibility of constructing an effective field theory applicable to the 
propagation of hard jets in soft dense matter.

The derivations in this effort have been restricted to parton propagation. The construction 
of an effective theory will require also the use of re-summed vertices. These will be 
derived in a subsequent publication when radiative processes will be dealt with. 
The inclusion of such contributions will allow for an all-twist calculation of radiative 
energy loss and comparisons with the energy loss formalism of Ref.~\cite{AMY}. 
The inclusion of radiative energy loss in an all twist formalism will allow, for the 
first time, a means to study the evolution of virtuality of the hard partons as they 
originate in hard collisions, traverse dense matter, escape and fragment into hadrons.

Confrontation with experiment~\cite{highpt} will require a convolution of the above formalism 
with both single~\cite{HT}  and multi-hadron~\cite{maj04a,maj04d} fragmentation functions.  
Even without the inclusion of energy loss, the above formalism may be directly applied to 
the measured transverse broadening of hard jets in cold nuclear matter as measured 
by the HERMES experiment at DESY~\cite{Airapetian:2000ks}. Such comparisons will be carried out 
in an upcoming publication. 

%%%%%%%%%%%%%%%%%%%%%%%%%%%%%%%%%%%%%%%%%%%%%%%%%%%%%%%%%%
%%%%%%%%%%%%%%%%%%%%%%%%%%%%%%%%%%%%%%%%%%%%%%%%%%%%%%%%%%
%%%%%%%%%%%%%%%%%%%%%%%%%%%%%%%%%%%%%%%%%%%%%%%%%%%%%%%%%%
%%%%%%%%%%%%%%%%%%%%%%%%%%%%%%%%%%%%%%%%%%%%%%%%%%%%%%%%%%
%%%%%%%%%%%%%%%%%%%%%%%%%%%%%%%%%%%%%%%%%%%%%%%%%%%%%%%%%%

\section{Acknowledgments}

%%%%%%%%%%%%%%%%%%%%%%%%%%%%%%%%%%%%%%%%%%%%%%%%%%%%%%%%%%
%%%%%%%%%%%%%%%%%%%%%%%%%%%%%%%%%%%%%%%%%%%%%%%%%%%%%%%%%%
%%%%%%%%%%%%%%%%%%%%%%%%%%%%%%%%%%%%%%%%%%%%%%%%%%%%%%%%%%
%%%%%%%%%%%%%%%%%%%%%%%%%%%%%%%%%%%%%%%%%%%%%%%%%%%%%%%%%%
%%%%%%%%%%%%%%%%%%%%%%%%%%%%%%%%%%%%%%%%%%%%%%%%%%%%%%%%%%

The authors wish to thank R.~J.~Fries and X.-N.~Wang for helpful discussions. 
This work was supported in part by the U.S. Department of Energy
under grant DE-FG02-05ER41367. 

%%%%%%%%%%%%%%%%%%%%%%%%%%%%%%%%%%%%%%%%%%%%%%%%%%%%%%%%%%
%%%%%%%%%%%%%%%%%%%%%%%%%%%%%%%%%%%%%%%%%%%%%%%%%%%%%%%%%%
%%%%%%%%%%%%%%%%%%%%%%%%%%%%%%%%%%%%%%%%%%%%%%%%%%%%%%%%%%
%%%%%%%%%%%%%%%%%%%%%%%%%%%%%%%%%%%%%%%%%%%%%%%%%%%%%%%%%%
%%%%%%%%%%%%%%%%%%%%%%%%%%%%%%%%%%%%%%%%%%%%%%%%%%%%%%%%%%

\appendix

%%%%%%%%%%%%%%%%%%%%%%%%%%%%%%%%%%%%%%%%%%%%%%%%%%%%%%%%%%
%%%%%%%%%%%%%%%%%%%%%%%%%%%%%%%%%%%%%%%%%%%%%%%%%%%%%%%%%%
%%%%%%%%%%%%%%%%%%%%%%%%%%%%%%%%%%%%%%%%%%%%%%%%%%%%%%%%%%
%%%%%%%%%%%%%%%%%%%%%%%%%%%%%%%%%%%%%%%%%%%%%%%%%%%%%%%%%%
%%%%%%%%%%%%%%%%%%%%%%%%%%%%%%%%%%%%%%%%%%%%%%%%%%%%%%%%%%

%%%%%%%%%%%%%%%%%%%%%%%%%%%%%%%%%%%%%%%%%%%%%%%%%%%%%%%%%%
%%%%%%%%%%%%%%%%%%%%%%%%%%%%%%%%%%%%%%%%%%%%%%%%%%%%%%%%%%
%%%%%%%%%%%%%%%%%%%%%%%%%%%%%%%%%%%%%%%%%%%%%%%%%%%%%%%%%%
%%%%%%%%%%%%%%%%%%%%%%%%%%%%%%%%%%%%%%%%%%%%%%%%%%%%%%%%%%
%%%%%%%%%%%%%%%%%%%%%%%%%%%%%%%%%%%%%%%%%%%%%%%%%%%%%%%%%%

\section{Gluon fusion diagrams}

%%%%%%%%%%%%%%%%%%%%%%%%%%%%%%%%%%%%%%%%%%%%%%%%%%%%%%%%%%
%%%%%%%%%%%%%%%%%%%%%%%%%%%%%%%%%%%%%%%%%%%%%%%%%%%%%%%%%%
%%%%%%%%%%%%%%%%%%%%%%%%%%%%%%%%%%%%%%%%%%%%%%%%%%%%%%%%%%
%%%%%%%%%%%%%%%%%%%%%%%%%%%%%%%%%%%%%%%%%%%%%%%%%%%%%%%%%%
%%%%%%%%%%%%%%%%%%%%%%%%%%%%%%%%%%%%%%%%%%%%%%%%%%%%%%%%%%

In this appendix, we briefly sketch the calculation of diagrams that include gluon fusion contributions 
such as in the right panel of Fig.~\ref{three_gluon}. Such diagrams originate from the inclusion of 
interaction terms from the QCD Lagrangian density, of the form, 

\bea
 \frac{-g}{2}  \left( \prt_\A A^a_\B (y_g) - \prt_\A A^a_\B (y_g)  \right) 
f^{abc} {A^a}^\A (y_g) {A^b}^\B (y_g) .
\eea

\nt
A term in the expansion of the hadronic tensor that would contain a fusion contribution will 
have the general form [ignoring ($-$) signs and factors of $i$ and $g$], 

\bea
W_{3g}^{\mu \nu} &=& \int dy dy_q dy_g  \lc J^\mu(y) \psibar (y_p) \g^\g t^d A^d_\g(y_p) \psi(y_p) \label{apndx_2}
\\
\ata \!\!\!\!\left( \prt_\A A^a_\B (y_g) - \prt_\A A^a_\B (y_g)  \right) 
f^{abc} {A^a}^\A (y_g) {A^b}^\B (y_g) J^{\nu} \rc .\nn
\eea

\nt
In the above equation, $J^{\mu}$  and $J^\nu$ are the currents at the locations $y$ and $0$ which 
couple to the photon and there is a contraction between the gluon vector potential appearing in the first 
line with any of the three gluon vector potentials in the second line. this contraction represents the 
one gluon propagator in the left panel of Fig.~\ref{three_gluon}. There are multiple reasons why 
such contributions are small. One reason is that unlike the ladder diagrams of Fig.~\ref{fig3}, 
the introduction of an extra coupling constant is not accompanied with the introduction of an 
extra length integral. As the momentum in the one gluon propagator is set by requiring that the 
two quark lines it connects to go nearly on shell, the gluon propagator cannot itself go on 
shell and as a result there is no length enhancement from the gluon propagator.

This contribution is also small due to color confinement and the condition that the forward 
momenta of all partons originating in the nucleus are not small enough to be in the saturation 
regime. The contraction of one of the vector potentials located at $y_g$ 
[in the second line of Eq.~\eqref{apndx_2}] with that at $y_p$, 
requires that the remaining two vector potentials be contracted into the nuclear state. As both 
vector potentials are at the same location, they must be contracted into the same nucleon. 
However, due to Eq.~\eqref{glue_color_conf}, this requires both gluons to be in a color 
singlet state. The presence of the antisymmetric factor $f^{abc}$ renders such contributions 
vanishing. Diagrams with three gluon vertices produce non-vanishing contributions if another 
gluon (at location say $y^\p$) were to be contracted with the two gluons (which fuse into one gluon), 
into the same nucleon state. 
Such contributions automatically require that another space integral over $y^\p$ be restricted to 
be in close proximity to $y_g$.
Also, such contributions necessarily require the introduction of the expectation of 3-parton operators in a
nucleon state, which, as we have argued previously, are small compared to the expectation of 
2-parton operators outside the saturation regime. 

%%%%%%%%%%%%%%%%%%%%%%%%%%%%%%%%%%%%%%%%%%%%%%%%%%%%%%%%%%
%%%%%%%%%%%%%%%%%%%%%%%%%%%%%%%%%%%%%%%%%%%%%%%%%%%%%%%%%%
%%%%%%%%%%%%%%%%%%%%%%%%%%%%%%%%%%%%%%%%%%%%%%%%%%%%%%%%%%
%%%%%%%%%%%%%%%%%%%%%%%%%%%%%%%%%%%%%%%%%%%%%%%%%%%%%%%%%%
%%%%%%%%%%%%%%%%%%%%%%%%%%%%%%%%%%%%%%%%%%%%%%%%%%%%%%%%%%
\section{Non-central cut diagrams}
%%%%%%%%%%%%%%%%%%%%%%%%%%%%%%%%%%%%%%%%%%%%%%%%%%%%%%%%%%
%%%%%%%%%%%%%%%%%%%%%%%%%%%%%%%%%%%%%%%%%%%%%%%%%%%%%%%%%%
%%%%%%%%%%%%%%%%%%%%%%%%%%%%%%%%%%%%%%%%%%%%%%%%%%%%%%%%%%
%%%%%%%%%%%%%%%%%%%%%%%%%%%%%%%%%%%%%%%%%%%%%%%%%%%%%%%%%%
%%%%%%%%%%%%%%%%%%%%%%%%%%%%%%%%%%%%%%%%%%%%%%%%%%%%%%%%%%

In Sect.~IV, the derivation of the transverse momentum diffusion of a hard parton was isolated to the 
case of a central cut. In Eq.~\eqref{n_equal_nprime} and the ensuing discussion, this was identified 
as the dominant contribution to the differential hadronic 
tensor at  a given order. In this appendix, we focus on diagrams at order $n + n^\p = 2m$ in terms of 
the strong coupling constant $g$, where $n \neq n^\p$. In particular, we focus on the case where $n = m-1$ and 
$n^\p = m + 1$. The starting point of this analysis is Eq.~\eqref{W_mu_nu_simple}.  
As in Sect.~IV, the hard part is expanded in a Taylor series in transverse momentum as 
in Eq.~\eqref{2n_der}, resulting in the integrand (written without phase factors and light-cone 
time ordering $\h$-functions as), 

\bea
& & H(p^+,q^-,p_\perp, y) S(p_\perp, y) \nn \\
& \ra & \prod_{i=1}^{m-1} \prod_{j=1}^{ m+1 } \frac{ \prt }{ \prt p_\perp^i } 
\frac{ \prt }{ \prt p_\perp^j }\left. H( p^+, q^-,p^i_\perp, {p^\p}^i_\perp ) \right|_{p_\perp = 0} \nn \\
\ata p^i_\perp ( - {p^\p}^j_\perp ) A^+_{a_i} ( y_i^-, y_\perp^i ) A^+_{a^\p_j} ( {y^\p}^-_j , {y^\p}^j_\perp ) .
\eea

As in the case of the central cut, the $\h$-functions time order the gluon insertions in 
incremental order from the photon vertex to the cut line. Each occurrence of the factor 
${ e^{i p^i_\perp \x y^i_\perp } p^i_\perp}_\A  A^+(y^i)$ 
may be converted, using integration by parts, to 
$-i e^{i p^i_\perp \x y_\perp }  {\prt^i_\perp}_\A A^+(y^i) \simeq -i e^{i p_\perp \x y_\perp }  {F_\perp}_\A^+(y^i)$.
As a result we obtain an over all sign of $(-i)^{(m-1)} i^{(m+1)} = -1$ compared to the central cut.

\begin{figure}[htbp]
%\begin{center}
%  \epsfxsize 80mm
%\hspace{0cm}
\resizebox{2in}{2.5in}{\includegraphics[0in,0in][5in,6in]{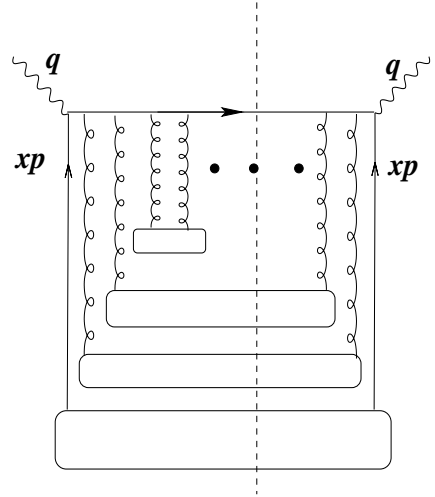}} 
%\vspace{0.25cm}
\caption{ An order $2m$,  contribution to $W^{\mu \nu}$ from a non-central cut}
    \label{fig6}
%  \end{center}
\end{figure}

The nuclear state is decomposed as before into a pseudo-free gas of nucleons and pairs of 
gluon insertions in singlet combinations is contracted into the $m$ chosen nucleons. Unlike the 
central cut, there is an excess of two gluon insertions on the left hand side of the cut as shown 
in Fig.~\ref{fig6}, where the rectangular blobs represent single nucleon states.
This extra pair is contracted into the same nucleon. 
For this specific example, the $m^{\rm{th}}$ and $(m+1)^{\rm{th}}$ gluon on the left hand side will be 
contracted into the same nucleon, as required for maximum length enhancement 
Contractions 
within the same nucleon, due to color confinement, tend to constrain the locations of the field operators 
$F^{\mu \nu}(y^-_m)$ and $F^{\mu \nu} (y^-_{m+1})$  to a region $y^-_{m+1} - y^-_{m} \leq y_{conf}$ the 
confining distance. Thus contractions between non-adjacent gluons, due to light-cone ordering $\h$-functions will constrain 
the length integrations of all the intermediate gluons to be within the size of a nucleon.

The transverse momentum integrations may 
now be carried out with the observation that in a very large nucleus, two point correlators only 
depend on the relative transverse distance between the two insertions. The final result is similar to 
that in Eq.~\eqref{two_gluon_cor} for $ i \leq m-1$ [see Eq.~\eqref{two_gluon_cor}].  
The resulting $\kd$-functions for the case of the $m^{\rm{th}}$ and $(m+1)^{\rm{th}}$ 
gluon which are both on the 
left of the cut, are slightly different, \tie,

\bea
& & \!\!\!\!\!\!\!\int d  \bar{y}^m_\perp  d \kd y^{(m+1)}_\perp e^{ - i  ( p^{m}_\perp +  p^{(m+1)}_\perp )\x \bar{y}^{m}_\perp  } 
e^{-i ( p^m_\perp  - p^{(m+1)}_\perp ) \x \kd y^m_\perp/2  } \nn \\
\ata \left<  t^{a_m} { F^{a_m}_\perp }^+_{\A_m}  (  \bar{y}^m_\perp + \kd y^m_\perp /2 )
t^{b_m} { F^{b_m}_\perp }^+_{\B_m}  (  \bar{y}^m_\perp - \kd y^m_\perp /2  )  \right> \nn \\
& = & (2 \pi)^2 \kd^2  \left(  p^m_\perp  + p^{(m+1)}_\perp \right) \int d \kd y^m_\perp
e^{i ( p^m_\perp )\x \kd y^m_\perp } \frac{- g_{\perp}^{\A_m \B_m}}{2} \nn \\
\ata \left< t^{a_m} { F^{a_m}_\perp }^+_{\rho_m}  (  \kd y^m_\perp /2 ) 
t^{b_m} { F^{b_m}_\perp }^{+ \rho_m}  (   - \kd y^m_\perp /2  )  \right> .
\eea

\nt
The transverse momentum $\kd$-function in the above equation 
imposes the condition that the transverse momentum brought in by one 
gluon is taken out by the other. Thus a pair of adjacent gluon insertions, in the  
limit of  transverse position invariance, infuses no net transverse momentum, into 
the final cut line. 

Incorporating the $m$ transverse momentum delta functions, leads to the following 
simplified expression for the hard part of the integrand, 

\bea
H &=& \left\{ \nabla^2_{l_\perp}\right\}^{(m-1)} 
\left[ \kd^2  \left( \vec{l} - \sum_{i=0}^{m-1} \vp^{\,\,i}_\perp  \right) \right] \nn \\
\ata \nabla^2_{p^m_\perp}  e^{i x^m_D p^+ \kd y_m^-} 
\eea
\nt
where, 
$\kd y_m^- = y_{m}^- - y_{(m+1)}^-$.

One notes the absence of the two transverse momenta $p^m_\perp, p^{(m+1)}_\perp$ from the argument 
of the $\kd$-function which determines the momenta of the cut-line ($l_\perp$). This will also be the case if 
the extra gluons were instead considered on the right hand side of the cut line. The essential point is that in the
case of translational invariance in a large nucleus, the leading length enhanced contribution from double scattering 
off a single nucleon produces no net transverse momentum broadening. As a result, such diagrams contribute to 
terms with a lower number of transverse momentum derivatives. The transverse momentum derivatives that act 
on the phase factors lead to suppressed contributions. 
An evaluation of the double derivatives on the phase factors, in the limit of vanishing transverse momenta, 
will leave an over all factor of $i \kd y_m^-/(q^-)$. As $\kd y_m^-$ is restricted to the size of the nucleon and 
the entire contribution is suppressed by a factor of $q^-$, such terms are parametrically suppressed in the 
evaluation of the transverse 
momentum distribution. Such terms are however important in terms of unitary  corrections to the total 
integrated cross section as will be demonstrated in the next appendix. \\

%%%%%%%%%%%%%%%%%%%%%%%%%%%%%%%%%%%%%%%%%%%%%%%%%%%%%%%%%%
%%%%%%%%%%%%%%%%%%%%%%%%%%%%%%%%%%%%%%%%%%%%%%%%%%%%%%%%%%
%%%%%%%%%%%%%%%%%%%%%%%%%%%%%%%%%%%%%%%%%%%%%%%%%%%%%%%%%%
%%%%%%%%%%%%%%%%%%%%%%%%%%%%%%%%%%%%%%%%%%%%%%%%%%%%%%%%%%
%%%%%%%%%%%%%%%%%%%%%%%%%%%%%%%%%%%%%%%%%%%%%%%%%%%%%%%%%%

\section{Unitarity corrections}

%%%%%%%%%%%%%%%%%%%%%%%%%%%%%%%%%%%%%%%%%%%%%%%%%%%%%%%%%%
%%%%%%%%%%%%%%%%%%%%%%%%%%%%%%%%%%%%%%%%%%%%%%%%%%%%%%%%%%
%%%%%%%%%%%%%%%%%%%%%%%%%%%%%%%%%%%%%%%%%%%%%%%%%%%%%%%%%%
%%%%%%%%%%%%%%%%%%%%%%%%%%%%%%%%%%%%%%%%%%%%%%%%%%%%%%%%%%
%%%%%%%%%%%%%%%%%%%%%%%%%%%%%%%%%%%%%%%%%%%%%%%%%%%%%%%%%%

In the evaluation of the transverse momentum distribution [$\phi(L^-,\vl_\perp)$] in 
Eq.~\eqref{solution}, as given by the solution of the diffusion 
equation [Eq.~\eqref{diffusion_eqn}], the focus was limited to the central cut diagrams as in Fig.~\ref{fig3}. It was 
argued that diagrams with non-central cuts do not contribute to the leading behavior of the transverse momentum 
distribution. These diagrams, however, play an important role in the unitarization of the integrated cross-section. 
The unitarity requirement states that 

\bea
 \frac{\prt \int d^2 l_\perp \phi(L^-,\vl_\perp)  } {\prt L^-} = 0. \label{unitarity}
\eea

\nt
One notes that the above requirement is satisfied by Eq.~\eqref{solution}. The unitarity requirement was 
tacitly assumed in formulating the overall normalization factor of the Gaussian distribution in Eq.~\eqref{solution}.
In what follows, we discuss the origin of such a constraint. 

The starting point is that of Eq.~\eqref{W_mu_nu_simple}. In the interest of simplicity we focus on the case of $n+n^\p=2$.
Cases with a higher number of gluon insertions follow a similar, albeit a more complicated pattern. For the 
case of two gluon insertion there are three cuts: the central cut with $n=n^\p = 1$, the left cut with $n=0, n^\p=2$ and the 
right cut with $n=2,n^\p=0$. The contribution to the central cut may be immediately evaluated from 
Eq.~\eqref{W_mu_nu_simple}, by setting $n=n^\p=1$. Integrating out the transverse momentum of the cut-line $l_\perp$
we obtain the hadronic tensor as 

\bea
W^{\mu \nu} &=& 
\int   \prod_{i=0}^1 d y_i^- d^2 y_\perp^i 
\prod_{j=1}^{1} d {y^\p}_j^- d^2 {y^\p}^j_\perp \nn \\
& & \int \prod_{i=0}^1 \frac{ d^2 p^i_\perp}{(2\pi)^2} 
\frac{d^2 {p^\p}^0_\perp} { (2\pi)^2}  \nn \\
\ata \hf \left( g^{\mu - } g^{\nu + } + g^{\mu +} g^{\nu -} - g^{\mu \nu}  \right) \nn \\
\ata  e^{-ix_B p^+ y^-} \prod_{i=0}^1 e^{-ix_D^i p^+ y_i^- } 
e^{i p^i_\perp \x  y^i_\perp  } \nn \\ 
\ata \prod_{j=0}^{1} e^{i {x^\p}_D^j p^+ {y^\p}_j^- } 
e^{-i{p^\p}^j_\perp \x  {y^\p}^j_\perp  }
\h ( y_1^-   - y^- ) 
\h ( {y^\p}_1^- ) \nn \\
\ata \lc A; p | \psibar(y^-,y_\perp) \g^+ \psi(0) 
\tr \left[  t^{a}  A_{a}^+ (y_1^-,y^1_\perp) \right. \nn \\
\ata \left.  t^{b} A_{b}^+ ( {y^\p}_1^-, {y^\p}^1_\perp ) \right]  | A;p \rc.
\label{W_mu_nu_2_glue}
\eea

\nt
The nuclear state is now decomposed into a gas of nucleon states with the quark 
wave function insertions restricted to one nucleon and the the gluon vector 
potential insertions restricted to another nucleon. 

In what follows we focus solely 
of the matrix element of  the two gluon insertions in a nucleon, which is referred to 
as the soft part $S$, 

\bea
S &=& \lc p |  A_{a}^+ ( y_1^-,y^1_\perp )  A_{b}^+ ( {y^\p}_1^-, {y^\p}^1_\perp ) | p \rc 
e^{i p^1_\perp \x  y^1_\perp  } e^{-i{p^\p}^1_\perp \x  {y^\p}^1_\perp  } \nn \\
&=& e^{i (p^1_\perp - {p^\p}^1_\perp ) \x \bar{y}_\perp  } e^{i (p^1_\perp + {p^\p}^1_\perp ) \x \kd y_\perp/2 } \nn \\
\ata \lc p |  A_{a}^+ ( \kd y_\perp/2 )  A_{b}^+ ( - \kd y_\perp/2 ) | p \rc, 
\eea
\nt
where, the transverse positions have been recast as $\bar{y}+\kd y/2$ and $\bar{y} - \kd y/2$.  
Transverse position invariance in a large nucleus is invoked in the last line of the above 
equation where the two point correlator is set to be independent of the mean transverse 
location $\bar{y}_\perp$.  The integration over the mean transverse location can be 
carried out to obtain the two dimensional delta function $\kd$-function $\kd^2 ( p^1_\perp - {p^\p}^1_\perp)$.
The hard part, which essentially consists of phase factors may now be expanded in 
a Taylor series in the transverse momenta $p_\perp, {p^\p}^1_\perp$ as in Eq.~\eqref{2n_der}. 
Carrying out the derivatives and setting the remaining transverse momenta to zero, gives the 
part that depends on the gluon insertions as

\bea
G &=& \int d y_1^-  d {y^\p}_1^- \h ( y_1^-   - y^- ) 
\h ( {y^\p}_1^- )  \nn \\
\ata i \frac{ (y_1^- - y_2^-)}{q^-}   \lc p |  F^+_\A  (y_1^-) F^{+ \A} ({y^\p}_1^-) | p \rc.
\eea
\nt
In the above equation, due to color confinement, $y_1^- - {y^\p}_1^-$ is always restricted to 
be smaller than the size of a nucleon. However, the mean location $( y_1^- + {y^\p}_1^-)/2$ 
is unrestricted and as a result this length integration provides an overall factor of $L^-$ and 
is the origin of the length enhancement of  the part of the cross section which corresponds to 
the central cut. 

 Setting $n=0,n^\p=2$ and $n=2,n^\p=0$, 
and following an almost similar method as above for the central cut yields the contributions 
from the left and right cuts. 
As shown in the preceding appendix, the terms originating from the left and right cuts provide almost identical contributions, except 
for an over all sign and the argument of the $\h$-functions. Relabeling the locations: \eg, $y_2 = y_1^\p$ in 
the case of the left cut (and similarly for the right cut to have the same set of position labels for all three cuts), yields 
the combination of $\h$-functions as

\bea
\Theta &=&  \h ( y_1^-   - y^- ) \h ( {y^\p}_1^- ) -  \h ( {y^\p}_1^-  - y_1^- ) \h ( y_1^-   - y^- ) \nn \\
&-& 
\h (  y_1^-  - {y^\p}_1^-) \h ({y^\p}_1^-). 
\eea 
\nt
As has been demonstrated earlier~\cite{HT,lqs,Fries:2002mu,Qiu:1990xx}, the above 
combination, along with the constraint from confinement destroys the length enhancement 
by restricting the average longitudinal location $( y_1^- + {y^\p}_1^-)/2$  to also lie within 
the size of a nucleon. The combination of all three cuts, in the evaluation of the integrated 
distribution, thus removes the overall factor of $L^-$ and as a result the integrated distribution 
obeys the condition of  Eq.~\eqref{unitarity}. The general case with $n$-gluon insertions may 
also be demonstrated to contain such unitarity corrections by a similar method of combining 
contributions from all cuts.

\end{document}